\documentclass[twocolumn]{aastex62}
\usepackage{lipsum}
\usepackage{afterpage}
\usepackage[utf8]{inputenc}
\usepackage{fourier} 
\usepackage{array}
\usepackage{makecell}
\usepackage{rotating}
\usepackage{array}
\usepackage{multirow}
\usepackage{graphics}
\usepackage[caption=false]{subfig}
\usepackage{enumitem}
\usepackage{fixltx2e}
\usepackage{mwe}
\usepackage{amsmath}
\usepackage{xcolor}
\usepackage[flushleft]{threeparttable}

\newcolumntype{L}[1]{>{\raggedright\let\newline\\\arraybackslash\hspace{0pt}}m{#1}}
\newcolumntype{C}[1]{>{\centering\let\newline\\\arraybackslash\hspace{0pt}}m{#1}}
\newcolumntype{R}[1]{>{\raggedleft\let\newline\\\arraybackslash\hspace{0pt}}m{#1}}

\graphicspath{{./}{figures/}}

\shorttitle{LONG-TERM VARIABILITY OF THE MILKY WAY CENTRAL SUPERMASSIVE BLACK HOLE}
\shortauthors{Chen et al.}

\begin{document}
\title{CONSISTENCY OF THE INFRARED VARIABILITY OF SGR A* OVER 22 YEARS}

\correspondingauthor{Zhuo Chen}
\email{zhuochen@astro.ucla.edu}

\author{Zhuo Chen}
\affiliation{Galactic Center Collaboration (GCC)}
\affiliation{Dept. of Physics and Astronomy, UCLA}

\author{E. Gallego-Cano}
\affiliation{Galactic Center Collaboration (GCC)}
\affiliation{Intituto de Astrofísica de Andalucía (CSIC), Glorieta de la Astronomía s/n, 18008 Granada, Spain}

\author{T. Do}
\affiliation{Galactic Center Collaboration (GCC)}
\affiliation{Dept. of Physics and Astronomy, UCLA}

\author{G. Witzel}
\affiliation{Galactic Center Collaboration (GCC)}
\affiliation{Dept. of Physics and Astronomy, UCLA}
\affiliation{Max Planck Institute for Radio Astronomy, Auf dem Hügel 69, D-53121 Bonn (Endenich), Germany}

\author{A. M. Ghez}
\affiliation{Galactic Center Collaboration (GCC)}
\affiliation{Dept. of Physics and Astronomy, UCLA}

\author{R. Sch\"{o}del}
\affiliation{Galactic Center Collaboration (GCC)}
\affiliation{Intituto de Astrofísica de Andalucía (CSIC), Glorieta de la Astronomía s/n, 18008 Granada, Spain}

\author{B. N. Sitarski}
\affiliation{Galactic Center Collaboration (GCC)}
\affil{Dept. of Physics and Astronomy, UCLA}
\affiliation{Giant Magellan Telescope Corporation, Pasadena, CA}

\author{E. E. Becklin}
\affiliation{Galactic Center Collaboration (GCC)}
\affiliation{Dept. of Physics and Astronomy, UCLA}

\author{J. Lu}
\affiliation{Galactic Center Collaboration (GCC)}
\affiliation{Dept. of Astronomy, UC Berkeley}

\author{M. R. Morris}
\affiliation{Galactic Center Collaboration (GCC)}
\affiliation{Dept. of Physics and Astronomy, UCLA}

\author{A. Dehghanfar}
\affiliation{Galactic Center Collaboration (GCC)}
\affiliation{Dept. of Physics and Astronomy, UCLA}
\affiliation{Institut de Planetologie et d’Astrophysique de Grenoble, \\
414 Rue de la Piscine, 38400 Saint-Martin-d’Heres, France}

\author{A. K. Gautam}
\affiliation{Galactic Center Collaboration (GCC)}
\affiliation{Dept. of Physics and Astronomy, UCLA}

\author{A. Hees}
\affiliation{Galactic Center Collaboration (GCC)}
\affiliation{Dept. of Physics and Astronomy, UCLA}
\affiliation{SYRTE, Observatoire de Paris, Universit\'e PSL, CNRS, Sorbonne Universit\'e,\\
LNE, 61 avenue de l'Observatoire 75014 Paris, France}

\author{M. W. Hosek, Jr.}
\affiliation{Galactic Center Collaboration (GCC)}
\affiliation{Dept. of Physics and Astronomy, UCLA}

\author{S. Jia}
\affiliation{Galactic Center Collaboration (GCC)}
\affiliation{Dept. of Astronomy, UC Berkeley}

\author{A. C. Mangian}
\affiliation{Galactic Center Collaboration (GCC)}
\affiliation{Dept. of Astronomy, University of Illinois, Urbana-Champaign}

\author{K. Matthews}
\affiliation{Galactic Center Collaboration (GCC)}
\affiliation{Division of Physics, Mathematics, and Astronomy, California Institute of Technology}

\submitjournal{ApJL}

\begin{abstract}

We report new infrared measurements of the supermassive black hole at the Galactic Center, Sgr A*, over a decade that was previously inaccessible at these wavelengths. This enables a variability study that addresses variability timescales that are ten times longer than earlier published studies. Sgr A* was initially detected in the near-infrared with adaptive optics observations in 2002. While earlier data exists in form of speckle imaging (1995 - 2005), Sgr A* was not detected in the initial analysis. Here, we improved our speckle holography analysis techniques. This has improved the sensitivity of the resulting speckle images by up to a factor of three. Sgr A* is now detectable in the majority of epochs covering 7 years. The brightness of Sgr A* in the speckle data has an average observed K magnitude of 16.0, which corresponds to a dereddened flux density of $3.4$ mJy. Furthermore, the flat power spectral density (PSD) of Sgr A* between $\sim$80 days and 7 years shows its uncorrelation in time beyond the proposed single power-law break of $\sim$245 minutes. We report that the brightness and its variability is consistent over 22 years. This analysis is based on simulations using \citet{Witzel et al. 2018} model to characterize infrared variability from 2006 to 2016. Finally, we note that the 2001 periapse of the extended, dusty object G1 had no apparent effect on the near-infrared emission from accretion flow onto Sgr A*. The result is consistent with G1 being a self-gravitating object rather than a disrupting gas cloud.

\end{abstract}

\section{Introduction} \label{sec:intro}
The Galactic Center, approximately 8 kpc \citep{Reid 1993} from Earth, is host to the closest known supermassive black hole (\citealt{Schodel et al. 2002,Ghez et al. 1998,Ghez et al. 2000,Ghez et al. 2005b,Ghez et al. 2008,Gillessen et al. 2009,Gillessen et al. 2017,Boehle et al. 2016}). This makes it an excellent laboratory for studying the accretion properties of supermassive black holes. The accretion flow onto the supermassive black hole at the Galactic Center gives rise to its radiative counterpart, Sgr A*, which appears to be very under-luminous compared to active galactic nuclei with comparable masses \citep{Melia Falcke 2001}. Several different theoretical models have been developed to describe Sgr A*'s accretion flow, including the well-known advection-dominated accretion flow model (ADAF, \citealt{Ichimaru(1977),Narayan Yi(1994),Narayan et al. 1995,Abramowicz et al. 1995}) and the radiatively inefficient accretion flow model (RIAF, \citealt{Yuan et al. 2003} ), both of which account for the low efficiency of the radiation loss of accreting gas and imply a hot and and geometrically thick accretion structure. 

An additional complexity and opportunity for modeling Sgr A* emission is that it is a variable source. Thus far the near-infrared (NIR) has proven to be a powerful window for characterizing Sgr A*'s variability \citep{Witzel et al. 2018}. Sgr A* was first detected in the NIR in 2002 with the first adaptive optics measurements of the Galactic Center (\citealt{Genzel et al. 2003,Ghez et al. 2004}). The NIR short-term variability of Sgr A* is well characterized as a red-noise process \citep{Press(1978)}. A power-law power spectral density (PSD) with a slope $\gamma_1 \approx$ 2 can describe the variability on short timescales of minutes to hours (\citealt{Do et al. 2009,Dodds-Eden et al. 2011,Witzel et al. 2012,Meyer et al. 2014,Hora et al. 2014}). \cite{Witzel et al. 2018} reported a break in the PSD at a timescale of $\tau_b = 245_{-61}^{+88}$ minutes, which constitutes the characteristic timescale of the variability process. The power law and break timescales have been a powerful way to study black hole accretion physics over a large range of luminosity and mass scales (e.g., \citealt{Meyer et al. 2008,Meyer et al. 2009,Do et al. 2009,Eckart et al. 2006b}). Thus far the longest NIR timescale that have been measured of Sgr A* is 1.9 years \citep{Meyer et al. 2009}.

Prior to and during the epoch when Adaptive Optics (AO) systems were coming online (2002 - 2005, VLT), the Galactic Center (GC) was studied at high angular resolution comparable to that achieved with AO at near-infrared wavelengths with speckle data from 1995 to 2005 at Keck. The initial analysis used the shift-and-add technique (SAA; \citealt{Eckart et al. 1995,Eckart Genzel 1996,Ghez et al. 1998,Ghez et al. 2000,Ghez et al. 2005a,Hornstein et al. 2002,Lu et al. 2005,Rafelski et al. 2007}). Sgr A* was not detected at this time owing to both the poorer sensitivity of these maps, which typically had detection limits ($<K_{lim}>$ = 15.7 mag, \citealt{Boehle et al. 2016}) comparable to or fainter than the average Sgr A* brightness ($<K_{Sgr A*, AO}>$ = 16.1 mag, see section \ref{sec:lightcurve}), and the short time baseline of observations, which allowed only limited knowledge of the orbits of nearby stars and the position of Sgr A* \citep{Hornstein et al. 2002}. Recently, the analysis of the speckle data has been improved with the speckle holography technique (\citealt{Primot et al. 1990,Schodel et al. 2013,Boehle et al. 2016}) to study two short-period stars, S0-38 and S0-2. This technique deepens the detection magnitude to $K < 17$ and opens the possibility of detecting Sgr A* over a much longer time baseline.

Detecting Sgr A* during the speckle era (1995 - 2005) also extends the time baseline for discrete accretion events searches. Of particular interest, is the spatially resolved, dusty source G1, which underwent a tidal interaction with the central black hole as it went through periapse in 2001 (\citealt{Sitarski et al. 2014,Pfuhl et al. 2015,Witzel et al. 2017}). This event may have increased the gas accretion onto Sgr A*. This object is similar observationally to G2, a cold, gaseous, highly-eccentric object orbiting Sgr A* that reached closest approach in early 2014 \citep{Gillessen et al. 2012}. G2 was originally posited to be a 3 Earth-mass pure gas cloud that would measurably impact the accretion flow and variability process as it was tidally sheared from the moment of periapse to $\sim 7$ years after periapse. However, no indication of this impact has so far been observed (e.g., \citealt{Witzel et al. 2014,Valencia-S. et al. 2015,Pfuhl et al. 2015,Hora et al. 2014,Witzel et al. 2017,Witzel et al. 2018}), but the interaction phase may extend a few years ($\sim$7 years or more) beyond periapse passage (e.g., \citealt{Pfuhl et al. 2015}). One hypothesis is that G1 and G2 are part of the same gas streamer \citep{Pfuhl et al. 2015}. If this is the case, then G1 may have also impacted the accretion flow and thereby caused an enhancement of accretion luminosity as it went through closest approach to Sgr A*, and perhaps a few years after. While the AO measurements only started in 2002 and did not cover the time baseline of G1’s periapse, the speckle datasets allow us to study whether G1 had any impact on the accretion flow related to its periapse passage. Moreover, two short period stars S0-2 and S0-16 went through the periapse (S0-16, 2000; S0-2, 2002) during our speckle era explored in this work and we can thus test whether they had any kind of impact on the variability of Sgr A*.

In this work, we further develop the speckle holography technique to analyze our speckle datasets (1995 - 2005). We make the first report of NIR detection of Sgr A* prior to 2002. The details of the ten years of data used in this work are described in section \ref{sec:observ}. Section \ref{sec:dataAna} presents the data analysis and methods, including the speckle holography image reconstruction and improvements, point sources extraction, photometric calibration and Sgr A* identification from speckle holography images. Section \ref{sec:res} presents the results of Sgr A* detections, observed brightness and its variability. Section \ref{sec:diss} discusses the impact of G1's periapse, and also simulations to explore how the variability of Sgr A* in the speckle data is compared to that at later times. We conclude with a summary in section \ref{sec:con} of the long-term activity of Sgr A* on timescales that are significantly longer than previous studies. Appendix \ref{sec:photometry} and Appendix \ref{sec:sourceanalyses} present details of photometry and source analyses used in this work for speckle holography images.

\section{Datasets} \label{sec:observ}
This paper is based on speckle imaging that was taken as part of the $Galactic$ $Center$ $Orbit$ $Initiative$ (GCOI) and that was originally presented in \citet{Ghez et al. 1998,Ghez et al. 2000,Ghez et al. 2005a,Ghez et al. 2008,Lu et al. 2005,Rafelski et al. 2007,Boehle et al. 2016}. From 1995 to 2005, the K[$2.2$ $\mu$m]-band speckle datasets of the Galaxy’s central $\sim 5'' \times 5''$ region were obtained with the W. M. Keck I 10m telescope and its near-infrared camera (NIRC; \citealt{Matthews,Matthews et al. 1996}). During each epoch, which combines observations ranging from 1-4 nights, roughly 10,000 short-exposure frames (t$_{exp}$ = 0.1 s) were obtained in data cubes consisting of 128 frames, which was the maximum number of frames that could be obtained in a single NIRC fits file. Within each cube, the time delay between the start time of each frame was 1.5 seconds in 1995 and 0.61 seconds thereafter. These series of short exposures were obtained with NIRC in its fine plate scale mode, with a scale of $20.396$ $\pm$ $0.042$ mas pixel$^{-1}$ and a corresponding field of view of $5.22'' \times 5.22''$. The data were obtained with the telescope in stationary mode, which keeps the pupil fixed with respect to the detector and causes the sky to rotate over a data cube. Our starting point for this paper's analysis is the individual frames that have had the instrumental effects removed (i.e., sky-subtracted, flat-fielded, bad-pixel-corrected, distortion-corrected) and that have been rotated to have a position angle of 0 degrees on the plane of the sky (see details in \citealt{Ghez et al. 1998,Ghez et al. 2000}). Table \ref{tab:speckle1} summarizes the 27 speckle observation epochs.

\section{Data Analysis and Methods} \label{sec:dataAna}

\subsection{Image reconstruction - A new implementation of Speckle Holography}\label{sec:speckleholography}

\begin{figure}
    \centering
    \includegraphics[width=3.4in]{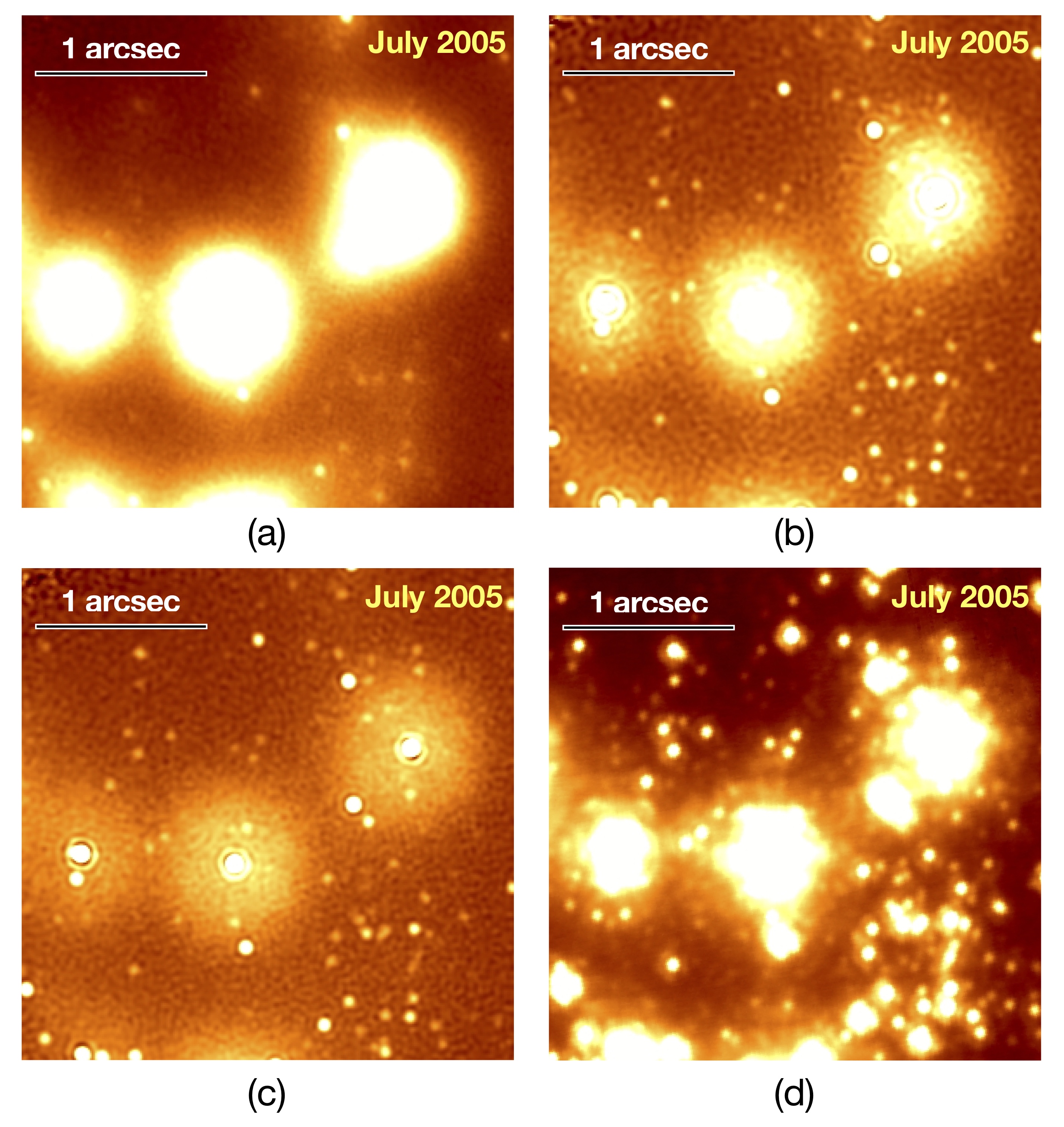}
    \caption{Comparison of the different high angular resolution techniques used to image the Galactic Center. The early data has been analyzed with (a) shift-and-add, (b) the original implementation of speckle holography (version 2\_1), (c) the new implementation presented in this work (version 2\_2). AO data taken at a similar time is shown in (d). The new speckle holography improves the sensitivity of the final images by up to a factor of three compared to the initial speckle holography.} 
    \label{fig:data}
\end{figure}

For this project, we have developed a new implementation of speckle holography. This builds on the work of \citet{Schodel et al. 2013}, which has been applied to the GCOI datasets presented in \citet{Meyer et al. 2012} (version 2\_0) and \citet{Boehle et al. 2016} (version 2\_1) to study the short-period stars. In theory, the speckle holography technique uses the instantaneous PSF, which is measured from a set of reference sources, to deconvolve, in Fourier space, the distorted images and realize the contribution of all speckle information to the final diffraction-limited core, as follows:
\begin{equation}
    \centering
    O = \frac{\langle I_mP_m^*\rangle}{\langle |P_m^2|\rangle}
\end{equation}
where O is the Fourier transform of the object, $I_m$ and $P_m$ are the Fourier transforms of the m-th short-exposure image and of its instantaneous PSF, respectively, and the brackets denote the mean over N frames. $P_m^*$ is the conjugate complex of $P_m$ \citep{Primot et al. 1990}.

In practice, Speckle holography images are constructed through an iterative process. A key component of this analysis uses the PSF fitting program StarFinder \citep{Diolaiti et al. 2000} (also see section \ref{sec:pointsources}). Below we detail the steps to construct the speckle holography images (version 2\_2), and how they differ from the implementation used in \citet{Boehle et al. 2016} (version 2\_1). 

\begin{enumerate}[noitemsep,nolistsep]
  \item Shift all short-exposure frames to align the brightest speckle of IRS 16C. Subtract a constant background, which is estimated for each individual frame, from each short-exposure frame for version 2\_2.
  \item Rebin the speckle frames from original 20 mas pixel$^{-1}$ scale down to 10 mas pixel$^{-1}$ scale. Bilinear and cubic interpolation are used in version 2\_1 and version 2\_2 respectively. 
  \item Combine and construct a shift-and-add (SAA) image from all data cubes per observing epoch. 
  \item Extract astrometry and photometry of stars in the SAA images with StarFinder to identify potential PSF reference stars for the speckle holography analysis. 
  \item Select the brightest isolated sources as PSF reference stars for speckle holography. Each data cube typically has 2-5 reference sources (IRS 16NE, IRS 16C, IRS 16NW, IRS 16SW, IRS 33N), depending on the centering and image quality of the data cube. 
  \item Estimate the instantaneous PSF for each speckle frame from the median of the aligned and flux-normalized images of the reference stars. For each PSF, we subtract a constant value of $b_g + n\times\sigma$ (n = 3 for speckle images), where $b_g$ is the background and $\sigma$ is the noise. All resulting negative values in the PSF are set to 0. As a final step, a circular mask is applied to the PSF and the PSF is normalized to a total flux of 1. In version 2\_2, we fixed a bug in StarFinder in which secondary stars that are not PSF reference stars were not being subtracted from the primary reference stars. 
  \item Improve the PSF estimate by subtracting all known secondary, contaminating sources near the reference stars in each frame, using the preliminary PSFs from step (6) and information from step (4). 
  \item Estimate the object Fourier transform O by applying equation (1). 
  \item Apodize O with a model for the optical transfer function (OTF) of the telescope. Here we use a Gaussian function for 10m aperture. 
  \item Reconstruct the image with inverse Fourier transform.
  \item Repeat of the process from step (4) with the holographically reconstructed image, which has significantly higher quality than the initial SAA image. 
  \item Create multiple images for error estimation. In version 2\_1, the dataset for each epoch is divided into 3 equal-quality subsets to produce 3 speckle holography maps. In version 2\_2, we use a bootstrapping method to produce 100 bootstrap datasets using sampling with replacement resulting in 100 speckle holography images produced from the same number of frames as the original dataset. 
\end{enumerate}
\vspace{3mm}

\begin{figure}
    \centering
    \includegraphics[width=3.4in]{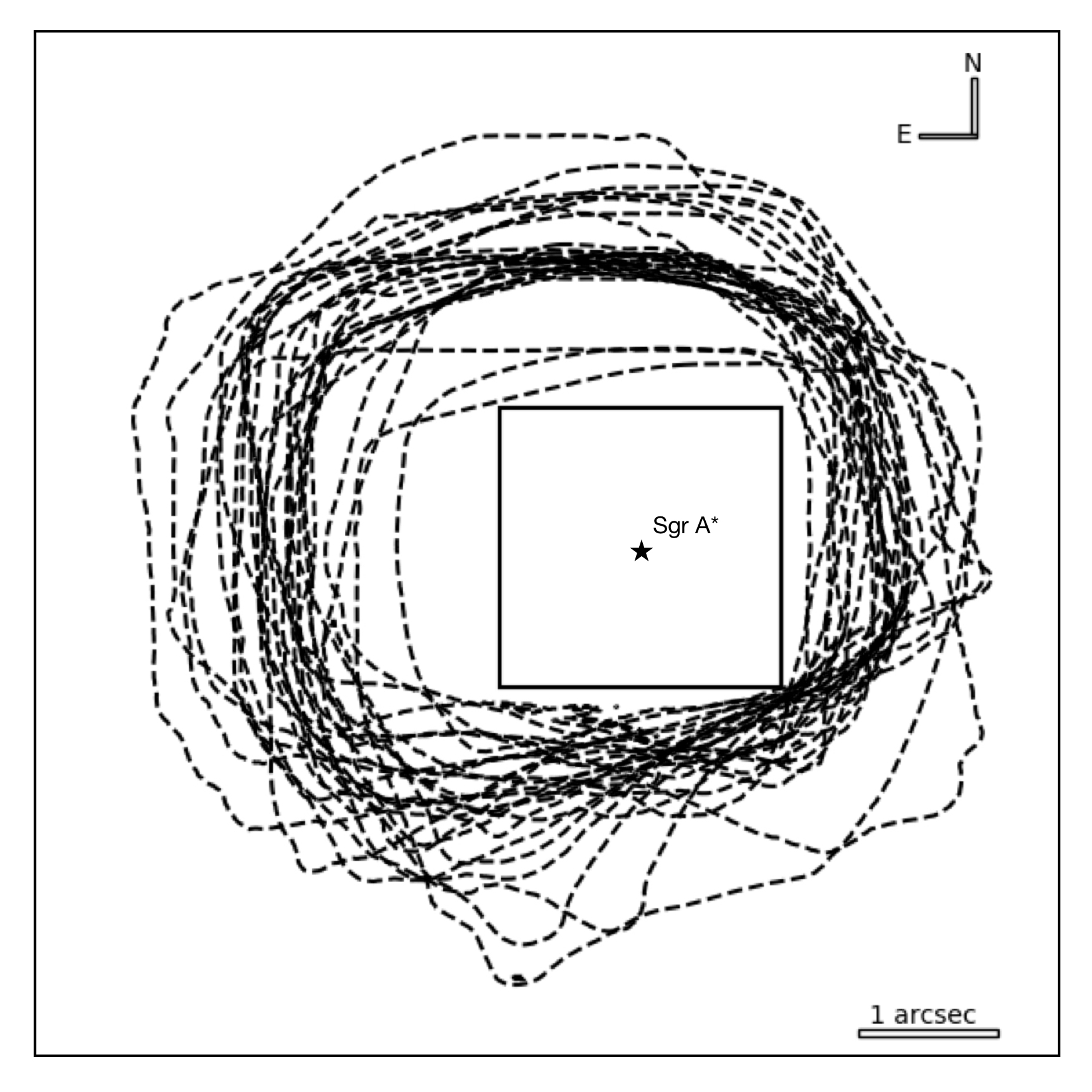}
    \caption{Comparison of the field of view for all the final speckle holography images. The dashed contours display the covered region for each epoch with contribution of over $80\%$ of the individual frames. In order to minimize the edge effects, this study considers only the central $2'' \times 2''$ region center around Sgr A* outlined with the solid line.} 
    \label{fig:image_nf}
\end{figure}

\begin{figure}
    \centering
    \includegraphics[width=3.5in]{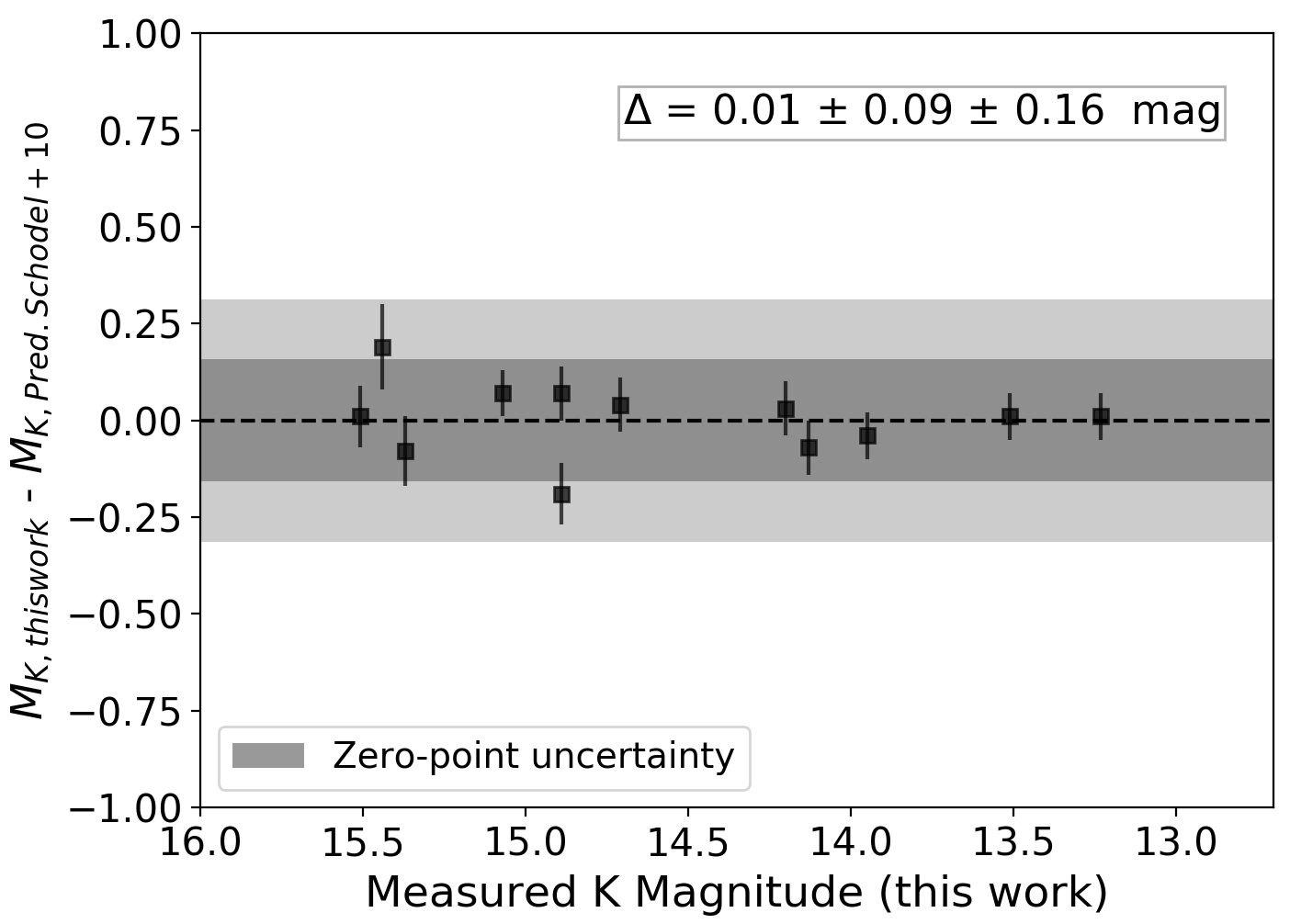}
    \caption{The difference between our measured K magnitude (from photometric system calibration) and the predicted K magnitude (bandpass corrected from \citet{Schodel et al. 2010} Ks) for calibrators used in \citet{Witzel et al. 2012}. The measured K magnitudes are consistent with the predicted ones, with an average difference of 0.01 $\pm$ 0.09 mag and a zero-point uncertainty $\overline{\sigma}_{zp}$ of 0.16 mags (dark band; and light band: 2$\overline{\sigma}_{zp}$). This verifies that the early IR measurements made based on speckle images are on the same photometric system as the later IR measurements obtained from AO.}
    \label{fig:k_ks}
\end{figure}

\begin{figure}
    \centering
    \includegraphics[width=3.4in]{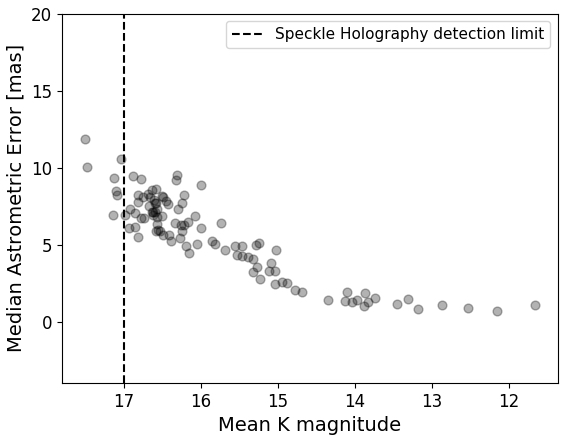}
    \caption{Astrometric uncertainty as a function of source brightness. While the brightest sources have an astrometric uncertainty of 1 mas, at the average speckle holography detection limit of K $\sim17$ mag, shown as the dashed line, the astrometric uncertainty is typically 10 mas.}
    \label{fig:searchradius}
\end{figure}

\begin{figure*}
\centering
\includegraphics[width=7in]{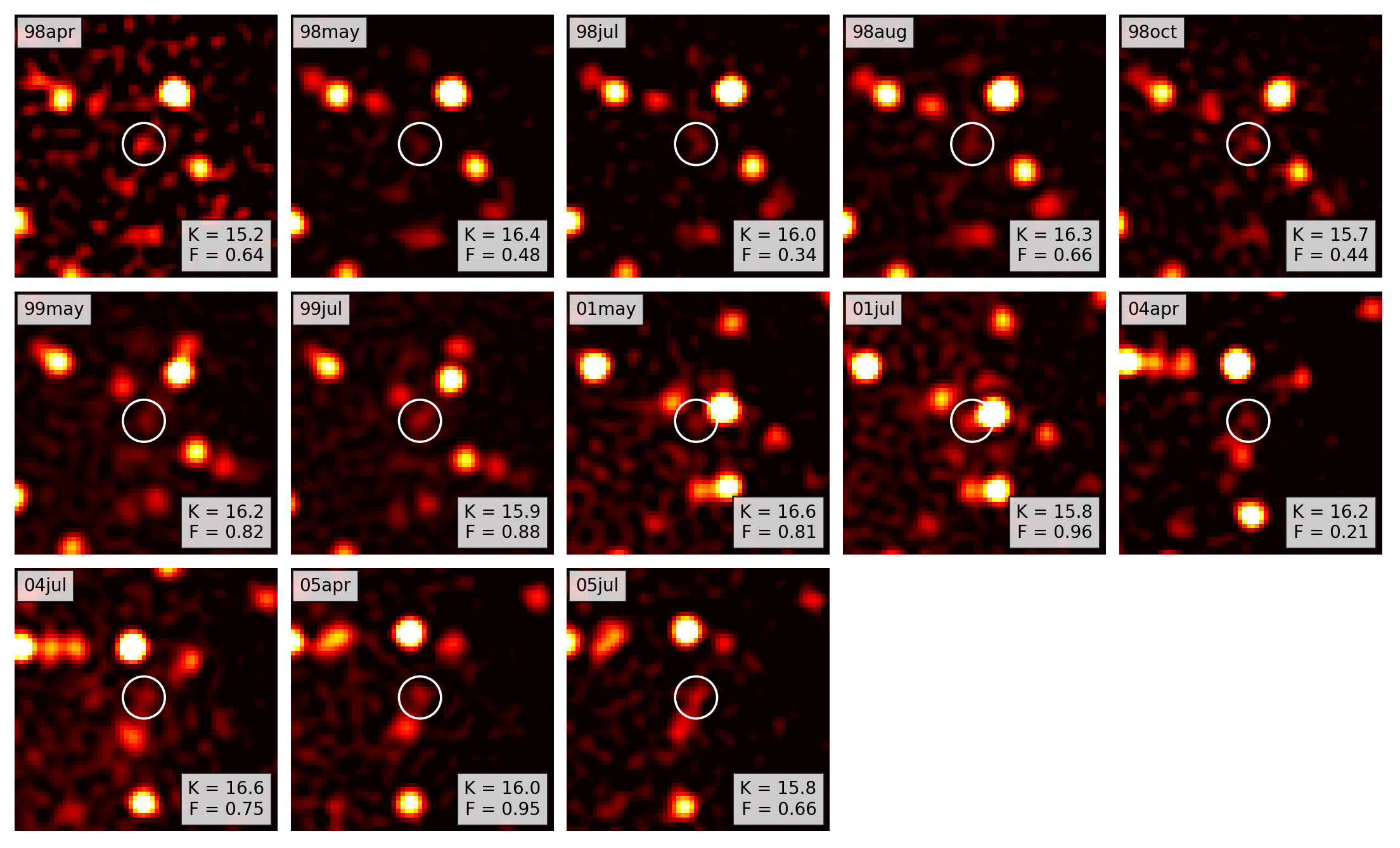}
\caption{Speckle holography images with new Sgr A* detections (white circles). For each detection, the K magnitude (K) and the bootstrap fraction (F) are provided. These are the first infrared detections of Sgr A* in the late 1990s and early 2000s. }
\label{fig:detections2}
\end{figure*}

\begin{figure*}
\centering
\includegraphics[width=6in]{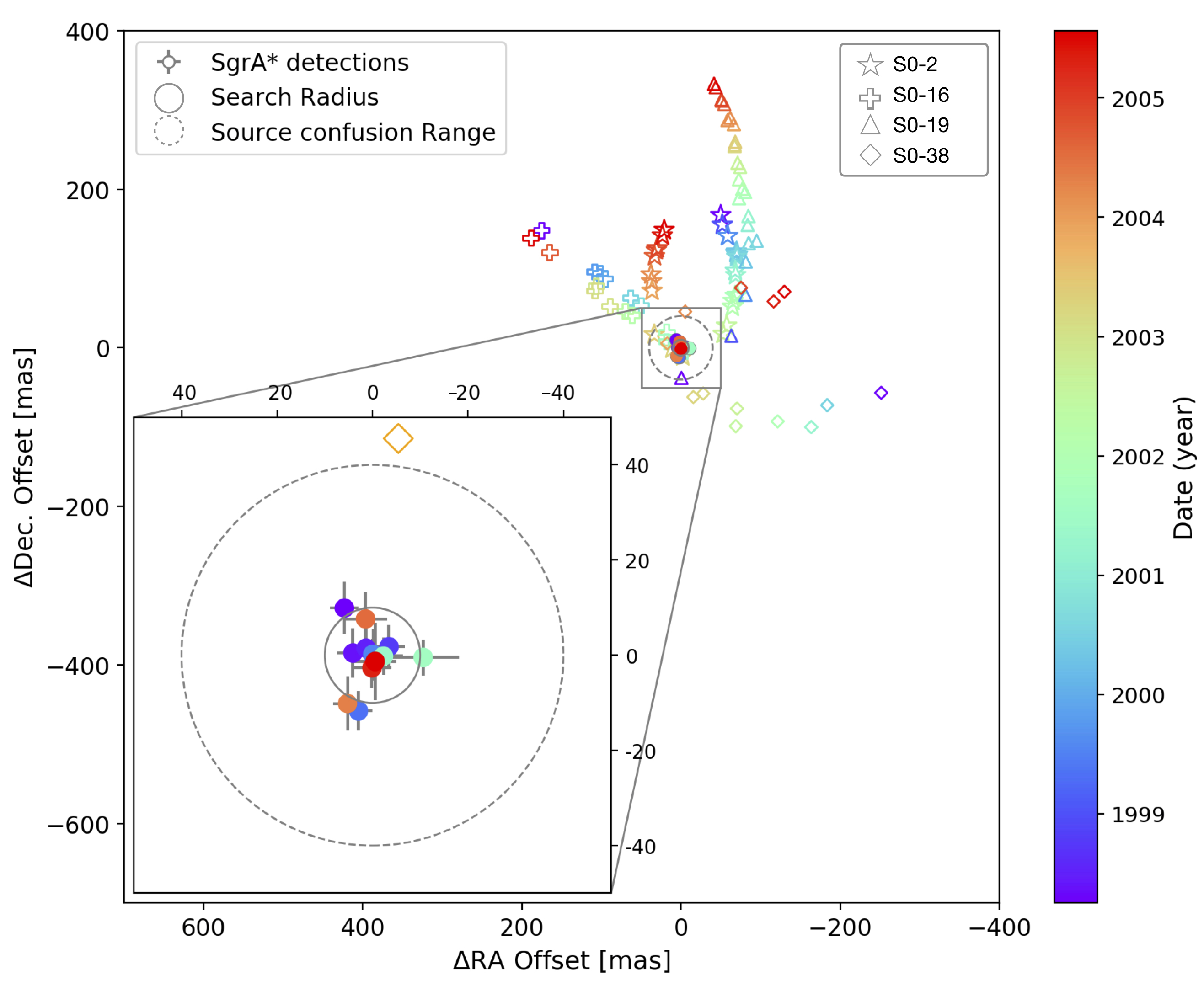}
\caption{RA and Dec. offsets of Sgr A* detections (points with error bars) and of nearby star detections (symbols marked with names) relative to the predicted position of Sgr A* ($(0, 0)$). Different colors show the corresponding epochs. The inset panel zooms into the central part of the region. The grey solid circle marks the search radius of 10 mas (see section \ref{sec:psffitting}) used to extract Sgr A* from the source list. The grey dashed circle shows the source confusion region (radius of 40 mas) within which nearby stars would cause bias and misdetection. We have removed confused Sgr A* detections and all nearby stars within this region in the inset panel. Overall, 13 detections of Sgr A* are free of source bias and used in this study.
}
\label{fig:confusion_2d}
\end{figure*}

The most important improvements of version 2\_2 compared to version 2\_1 are listed here:
\begin{itemize}[noitemsep,nolistsep]
  \item Subtracting a constant sky background from each frame results in less background variation in the combined images, and in significantly suppressed edge effects near the edge of the field of view (Step 1)
  \item Version 2\_1 only used IRS 16C as PSF reference source. Version 2\_2 uses up to 5 stars as PSF reference sources, depending on the instantaneous field of view (Step 5)
  \item The bootstrapping method results in a more robust estimate of the astrometric and photometric uncertainties (Step 12)
\end{itemize}

Figure \ref{fig:data} shows the central region of the final reconstructed image from 2005 July. Compared to the earlier implementation of speckle holography and SAA analysis, the new analysis (version 2\_2) has both improved the image quality and reduced the edge effect artifacts.

\subsection{Point sources extraction from speckle holography images}\label{sec:pointsources}

Point sources are extracted from each epoch's final reconstructed image using StarFinder. Here, like in section \ref{sec:speckleholography}, we use the version of StarFinder utilized in \citet{Boehle et al. 2016} setting the cross-correlation threshold to 0.8, and a slightly lower minimum Signal-to-noise ratio cut (3$\sigma$, vs. 5$\sigma$). In order to estimate the astrometric and photometric uncertainties for the sources extracted, we perform StarFinder on 100 bootstrap images for each epoch and then calculate the standard deviation. 

In order to minimize the impact of edge effects of speckle holography images, we restricted our analysis in the central $2'' \times 2''$ region center around Sgr A* for all epochs in the rest of this work. Owing to the observing strategy, in which stationary mode was used (see section \ref{sec:observ}), the final image has significant variations in the number of individual frames that contributes to the each pixel towards the outer edge of the image. See contours in Figure \ref{fig:image_nf}. 

Because speckle holography is a Fourier deconvolution technique, point-like artifacts can be produced in the middle of the field of view from edge effects, background and PSF extraction. We therefore require sources to be detected above a minimum threshold number of bootstrap images to be considered real. This threshold is set by demanding that the probability of fake detections be less than $1\%$ (see Appendix \ref{sec:lowerthreshold}). The bootstrap threshold for each epoch is reported in Table \ref{table:chartable} and with an average of $10\% \pm 7\%$ \footnote{We note that this threshold is lower than in other GCOI studies because we have strong prior knowledge of the source (Sgr A*) location.}.

\subsection{Photometric calibration}\label{sec:photometry_calib}
The list of extracted sources from speckle holography images is photometrically calibrated using a two-step procedure described in detail in the Appendix \ref{sec:photometry}. This process results in an average photometric systematic zero-point uncertainty of 0.14 mag and an average relative photometric uncertainty of 0.04 mag respectively in NIRC K bandpass (see Table \ref{tab:speckle1}). While we use the standard photometry in \citet{Blum et al. 1996} as the initial system calibration, we find that our photometry of stars in the galactic center is consistent with both the photometric systems of \citet{Witzel et al. 2012} and \citet{Schodel et al. 2010} for the AO measurements of Sgr A* obtained between 2004 and 2017. The overall photometric difference between \citet{Schodel et al. 2010} and us is $\sim$ 1 \% with an average difference of only $0.01 \pm 0.09$ mags and a zero-point uncertainty of 0.16 mags. See Figure \ref{fig:k_ks}.

\subsection{Comparison of speckle holography implementation}\label{sec:speckle_com}

This work has introduced an improved implementation of speckle holography. Appendix \ref{sec:v1v2com} compares in detail the performances between the current and the old versions and shows the clear improvements of version 2\_2. In particular, the version 2\_2 is 0.4 mag deeper on average but can be as much as a factor of three more sensitive in the extreme. From here on, we only consider the analysis based on the new speckle holography (version 2\_2), which has an average detection limit of 16.9 mag (see Table \ref{tab:speckle1}).

\subsection{Identification of Sgr A* from source list}\label{sec:psffitting}

We use the following steps to identify Sgr A* in the source list \footnote{Here we use the source list excluding the most likely artifact sources identified in Appendix \ref{sec:lowerthreshold}}:

\begin{enumerate}[noitemsep,nolistsep]
  \item Determine the position of Sgr A* in each epoch from the offsets between Sgr A* and IRS 16C and S0-2 since they are bright enough (IRS 16C: K = 9.8 mag; S0-2: K = 14.2 mag) to be always identified and obtained the accurate positions in the image. The offsets were generated by aligning all of our speckle holography and adaptive optics datasets together. See \citet{Ghez et al. 2008}, \citet{Gautam et al. 2018}, \citet{Jia et al. 2019}, and \citet{Sakai et al. 2019} for more details. 
  \item Search Sgr A* in the source list using the expected positions estimated in step (1). Sgr A* detected candidates are extracted if within the search radius of 10 mas. The search radius was determined by exploring the median astrometric error for all real detections in speckle epochs (Appendix \ref{sec:lowerthreshold}) in the central $2'' \times 2''$ region. See Figure \ref{fig:searchradius}. Empirically for sources with $K \sim 17$ mag (average speckle holography detection limit for all epochs, see Appendix \ref{sec:v1v2com}), the astrometric uncertainty is typically 10 mas. Based on this, we do not expect any real Sgr A* detections beyond 10 mas search radius. 
  \item Identify epochs where there is confusion with a known star that is passing within a 40 mas radius \citep{Jia et al. 2019} away from Sgr A*.
\end{enumerate}

\section{Results} \label{sec:res}
\subsection{Sgr A* detections}

Results on the detections of Sgr A* in our 27 epochs of speckle holography imaging data fall into 4 categories: detections without source confusion (13); non-detections (5); confusion with brighter sources (8); confusion with fainter sources (1). See Table \ref{table:chartable} for details.  

\subsubsection{Detections without source confusion}\label{sec:detections13}

Figure \ref{fig:detections2} presents the images of epochs with Sgr A* detections that are free from source confusion. The average observed magnitude of Sgr A* as obtained from these 13 epochs, which span the seven years period 1998 - 2005, is $K = 16.0$ $\pm$ $0.4$ (standard deviation, std) with average relative photometric uncertainty of 0.1 mag, corresponding to the average observed flux density of $0.35$ $\pm$ $0.13$ mJy with average relative photometric uncertainty of 0.04 mJy.

\subsubsection{Non-detections}\label{sec:nondetect}

Among the 18 available epochs without source confusion (see section \ref{sec:psffitting}), Sgr A* is not detected in 5 epochs. The brightness limit for Sgr A* in these epochs was determined by the source detection limit of that epoch (defined in Appendix \ref{sec:lowerthreshold}) and has values ranging from 16.0 to 17.0 mag.

\subsubsection{Detections confused with brighter sources}

In 8 epochs, Sgr A* was confused with a source brighter than its average value. As shown in Figure \ref{fig:confusion_2d} confusion occurred with the following brighter stars: S0-2, $K_{ave}$ = 13.6 mag in April, May, and July 2002; S0-19, $K_{ave}$ = 15.0 mag in June 1995; S0-16, $K_{ave}$ = 15.1 mag in April, May, July, and October 2000. These epochs were removed from further analysis.

\subsubsection{Detections confused with fainter sources}

In one epoch, Sgr A* was confused with a source fainter than its average value (see Figure \ref{fig:confusion_2d}). In April 2003, there is a source detection which is the combination of Sgr A* ($K_{ave}$ = 16.0 mag) and S0-38 ($K_{ave}$ = 16.5 mag). Because the confusing source is fainter, the constraints on Sgr A* can be obtained from the photometry ($K$ = 16.4 mag) and the astrometry ($K$ = 16.7 mag). These are comparable to the detection limit of this image ($K$ = 16.3 mag). We therefore place a limit on Sgr A* in this epoch of 16.3 mag.

\begin{figure}
\centering
\includegraphics[width=3.5in]{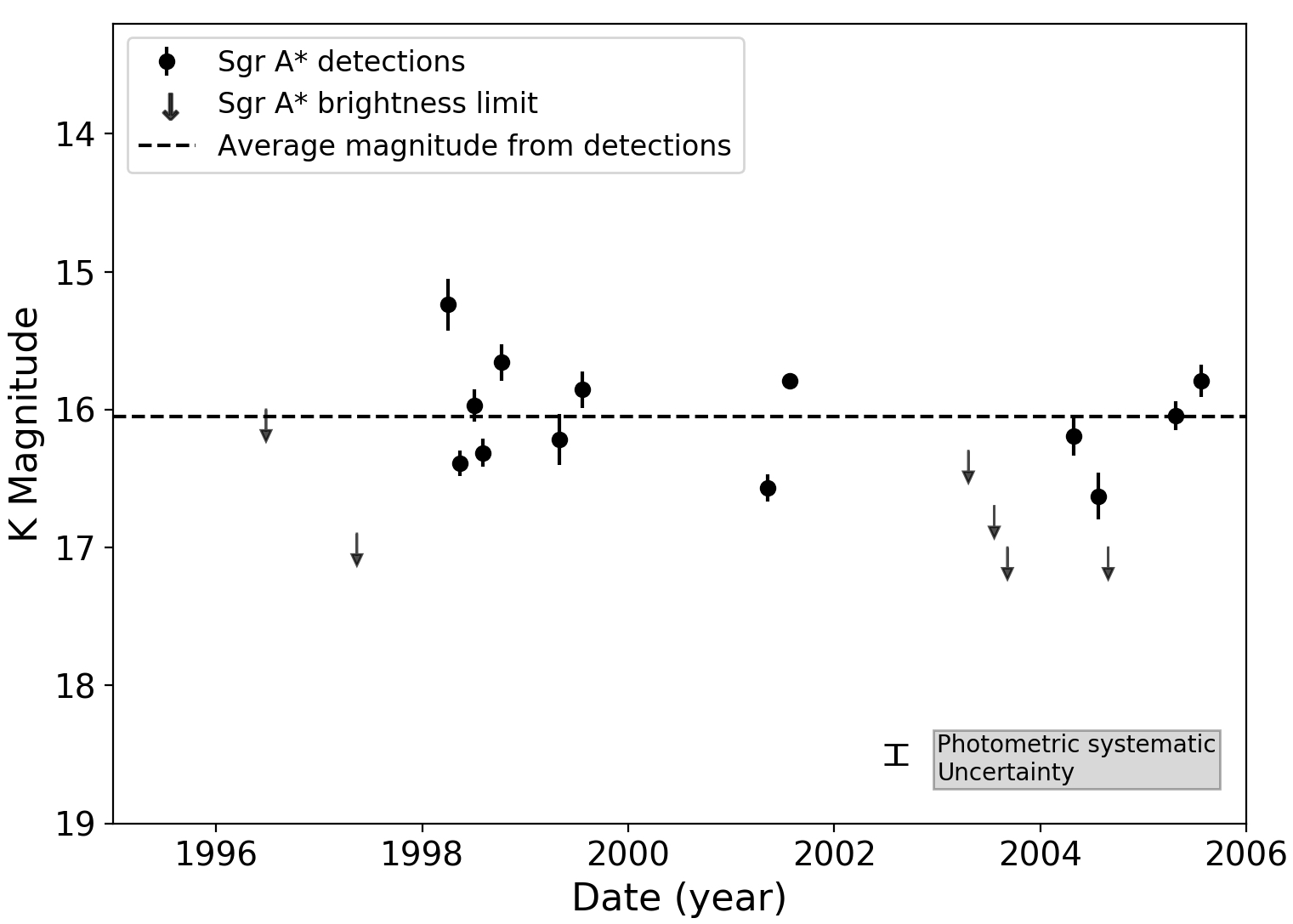}
\caption{Light curve of Sgr A* from 1998 to 2005. The points with error bars are the confirmed detections of Sgr A*. The arrows mark the brightness limit of Sgr A* in other non-detected epochs. The errorbar on the right bottom shows the average photometric systematic uncertainty of 0.14 mag. The average magnitude from the 13 detections is 16.0 mag. }

\label{fig:results}
\end{figure}

\begin{figure}
\centering
\includegraphics[width=3.5in]{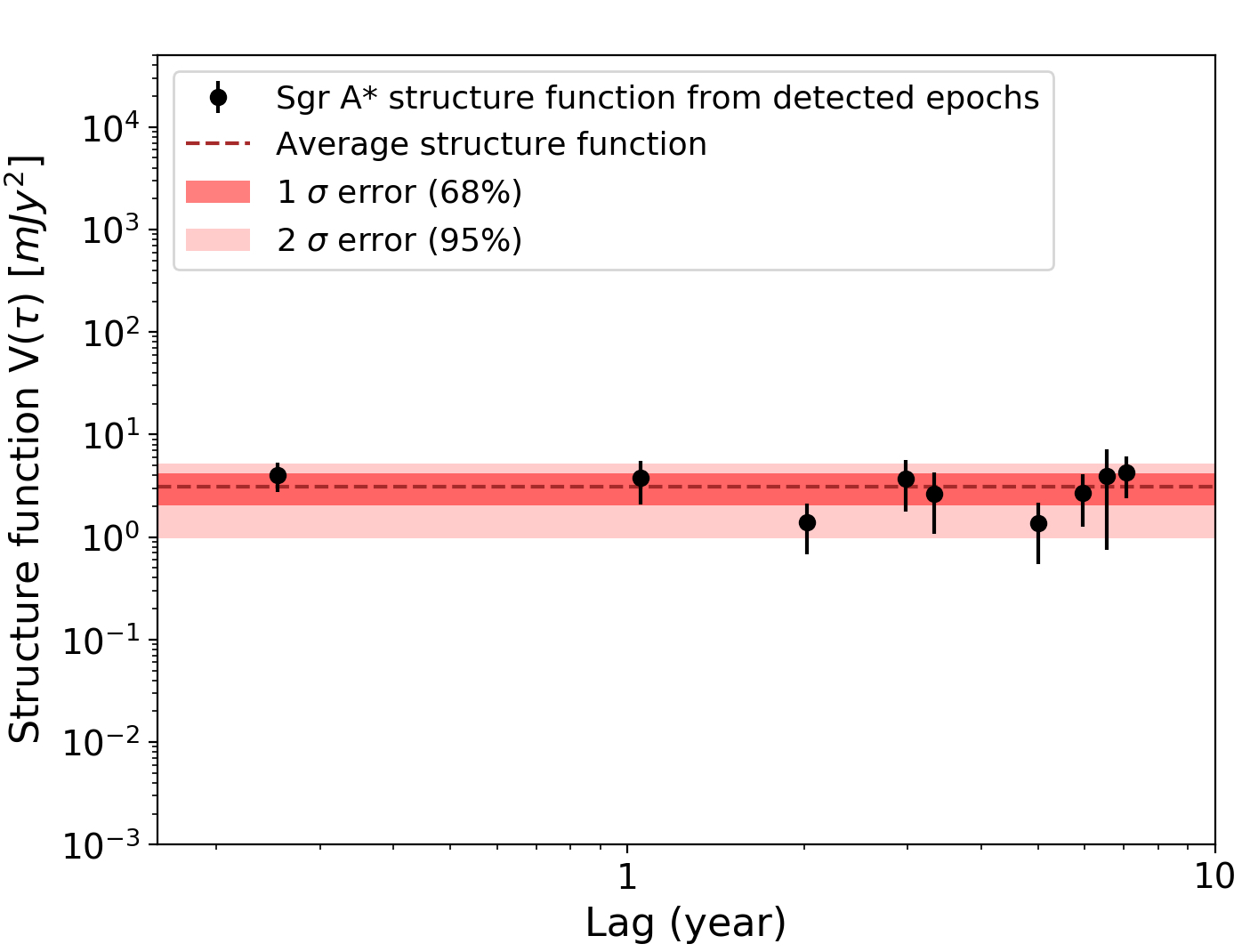}
\caption{Sgr A* structure function $V(\tau)$. The average structure function from 13 detections is 3.1 $\pm$ 0.3 $mJy^2$ (red band). }
\label{fig:sf}
\end{figure}

\begin{figure}
\centering
\includegraphics[width=3.5in]{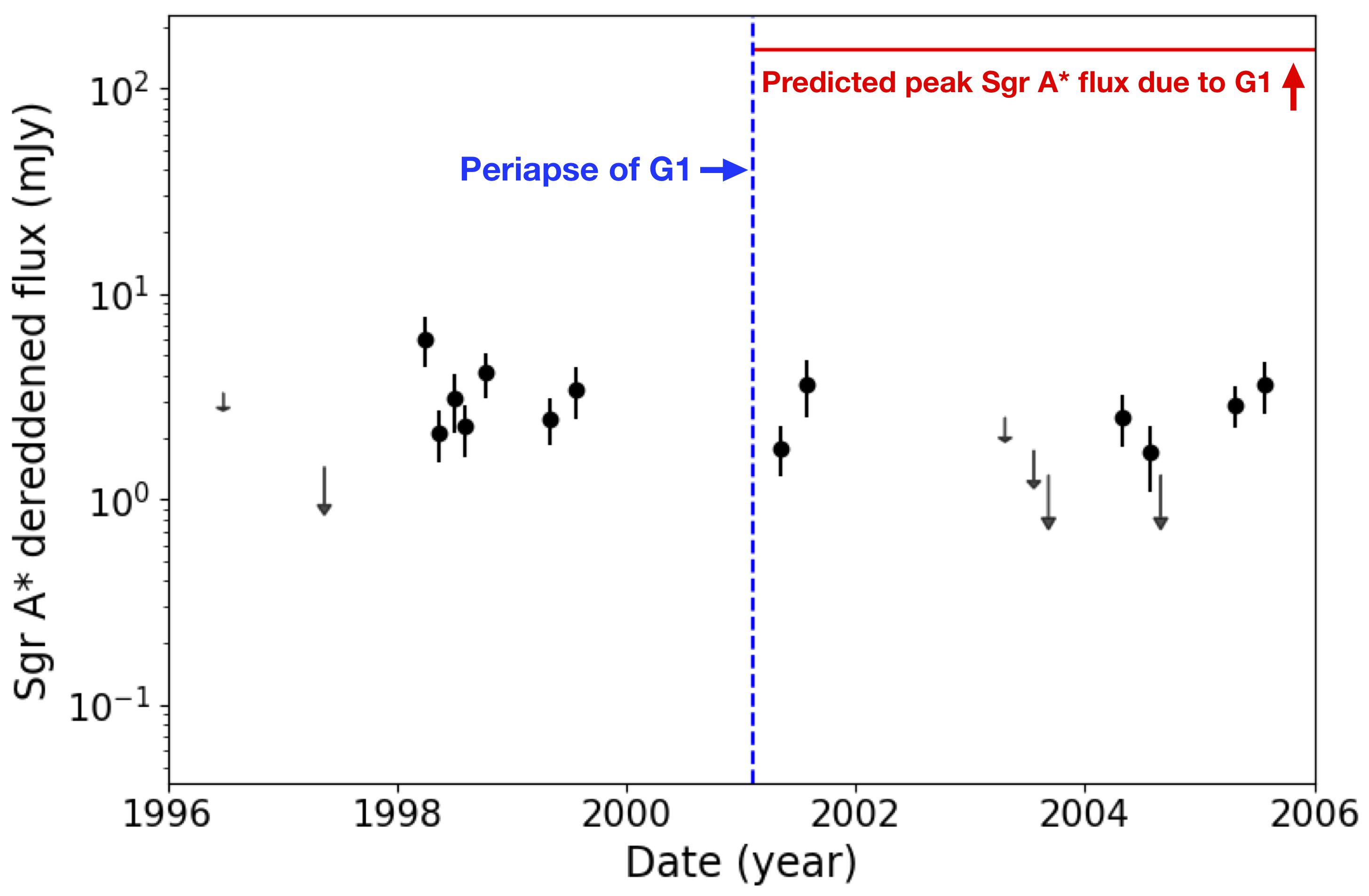}
\caption{Impact of G1's closest approach on the brightness of Sgr A*. The blue dashed line marks the G1's periapse (2001), and red line marks the predicted peak Sgr A* flux due to the closest approach of G1. No brightening or flares of Sgr A*, i.e. no apparent impacts of G1 on the Sgr A* infrared emission, were observed between 2001 and 2005.}

\label{fig:g1}
\end{figure}

\subsection{Sgr A* Brightness (1996 - 2005)}\label{sec:lightcurve}

\subsubsection{Average brightness}\label{sec:dered}
Figure \ref{fig:results} shows all our detections and detection limits of Sgr A* from 1996 to 2005. We convert the observed values into dereddened flux densities using the relationship $F_{K_s} = 6.67\times10^5\times 10.0^{-0.4\times (K_s - A_{ext,K_s})}$ mJy \citep{Tokunaga} and assuming $A_{ext,K_s} = 2.46$ mag extinction (\citealt{Schodel et al. 2010,Schodel et al. 2011}). Then we use the filter transformation $F_{K} = 1.29 F_{K_s}$, which is computed for the observed color of Sgr A* (H - K = 2.6 mag), to convert from K$_s$ fluxes to K fluxes (see Table \ref{tab:filtersgra}). See Appendix \ref{sec:calib} for more discussion of the filter transformations. Here in order to present the absolute dereddened fluxes, all uncertainties contain both the photometric systematic uncertainties and relative uncertainties (see Appendix \ref{sec:photometry}). The average detected dereddened flux density (ignore the brightness limit) is $3.4$ $\pm$ $1.2$ mJy (standard deviation) with average uncertainty of 0.6 mJy (0.4 mJy of relative photometric uncertainty only). If including the brightness limit, (a) treat brightness limit as a value: the upper limit of the dereddened flux density is $2.9$ mJy; (b) treat brightness limit as zero: the lower limit is $2.3$ mJy. The average from detections and the variance (see following section \ref{sec:sf_obs}) are consisent with the expectations from simulations based on more recent AO observations (2006 - 2017) as modeled from \cite{Witzel et al. 2018} (see section \ref{sec:simtoobs}, Figure \ref{fig:flux_ave} and Figure \ref{fig:struct_ave}).

\subsubsection{Long timescale Variance (40 days - 7 years)}\label{sec:sf_obs}

We used the first-order structure function to characterize the variability of Sgr A* over the 7 years. This approach is similar to the analysis of the timescale and the intrinsic variability of AGN light curves (e.g., \citealt{Simonetti et al. 1985,Hughes et al. 1992,Paltani 1999}) and Sgr A* short time variability (e.g., \citealt{Do et al. 2009,Witzel et al. 2012,Witzel et al. 2018}). For the set of flux measurements shown in the light curve, s(t), the first-order structure function V($\tau$) measures the flux density variance for a given time separation $\tau$: 
\begin{equation}
    \centering
    V(\tau) \equiv \langle[s(t+\tau) - s(t)]^2\rangle
\end{equation}
We calculated $[s(t+\tau) - s(t)]^2$ using Sgr A* dereddened fluxes reported in this work for all possible pairs of time lags from real speckle observational time series. Then we put the variances into bins with a bin size of 100,000 minutes ($\sim$ 70 days). This yields 9 bins covering the timescale between 42 days and 7.3 years, and each bin contains at least 5 and as many as 19 data points. In each bin, we assigned the median lag time to be the lag time for that bin, and the average of the V($\tau$) values to be the value of the structure function at that lag. The error of the structure function for each bin is calculated from $\sigma_{bin}/\sqrt{N_{bin}}$. Here $\sigma$ is the standard deviation of the V($\tau$) values and N is the number of points in that bin. The structure functions calculated with Sgr A* dereddened fluxes from observations are presented in Figure \ref{fig:sf}. The structure function is flat over timescale from 42 days to 7.3 years, and has an average value of 3.1 $mJy^2$ with the standard deviation of 1.1 $mJy^2$.

\section{Discussion} \label{sec:diss}

\subsection{Impact of G1 passage}
Based on 7 years of speckle holography data, we can use the variability of Sgr A* as the indication of the accretion activity in the past between 1998 and 2005. During this time, the dusty source G1 went through the closest approach in 2001. This object has similar observational properties to G2, which is the first example of a spatially resolved object tidally interacting with Sgr A*. The hypotheses for the nature of G1 are similar to those proposed to explain G2. These predictions range from compact gas clouds to the product of binary mergers (\citealt{Gillessen et al. 2012,Phifer et al. 2013,Murray-Clay et al. 2012}). Predictions of compact gas clouds near Sgr A* suggest that they may increase the accretion flow onto Sgr A*. For G2, one prediction \citep{Schartmann et al. 2012} is that if it is a gas cloud, it may be tidally disrupted and accrete onto Sgr A*, increasing the black hole's luminosity by up to a factor of 80. Since G1’s tidal radius is even smaller than that of G2, it would be more influenced by the black hole. If G1 was a gas cloud and some of the gas had been accreted onto the black hole, we may expect the additional accretion at the periapse. We marked the time following of G1’s periapse passage in Figure \ref{fig:g1}. Between 2001 and 2005, there is no increase in flux observed in Sgr A*. Sgr A* was quite steady with no evidence of large variations in brightness. The result is consistent with G1 being a self-gravitating object (such as a merger of two stars) as suggeted by \citet{Witzel et al. 2017}.

\begin{figure}
\centering
\includegraphics[width=3.5in]{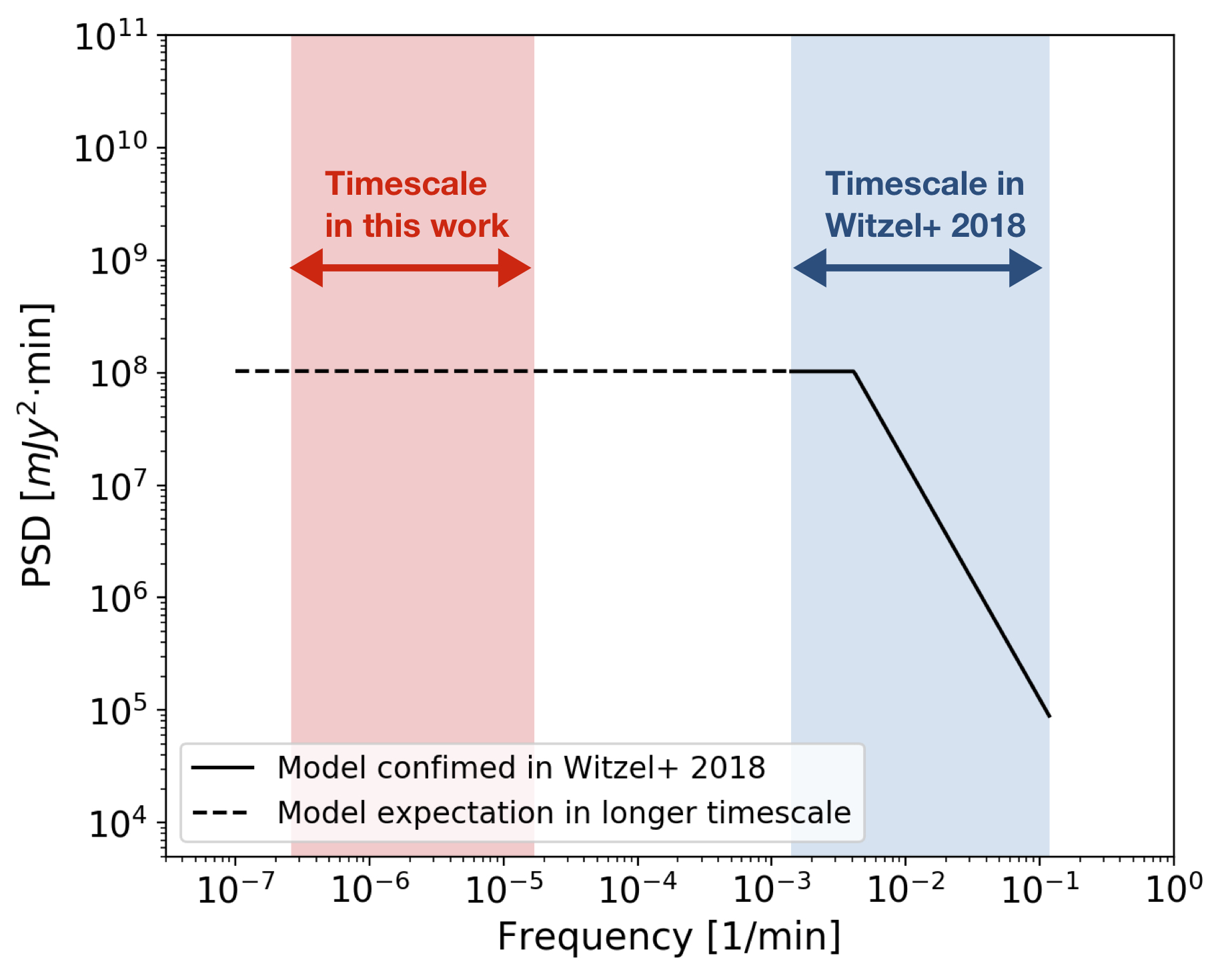}
\caption{The modeled power spectrum density (PSD) of Sgr A* presented in \citet{Witzel et al. 2018}. The solid line shows the broken power law PSD which has been confirmed in previous short-term variability observations with a break timescale of $\tau \sim$ 245 minutes. The dashed line shows the assumed flat PSD in longer timescale. The blue band marks the timescale probed in existed studies, and the red band marks the timescale that we explored in this work.}
\label{fig:psd_model}
\end{figure}

\begin{figure}
\centering
\includegraphics[width=3.4in]{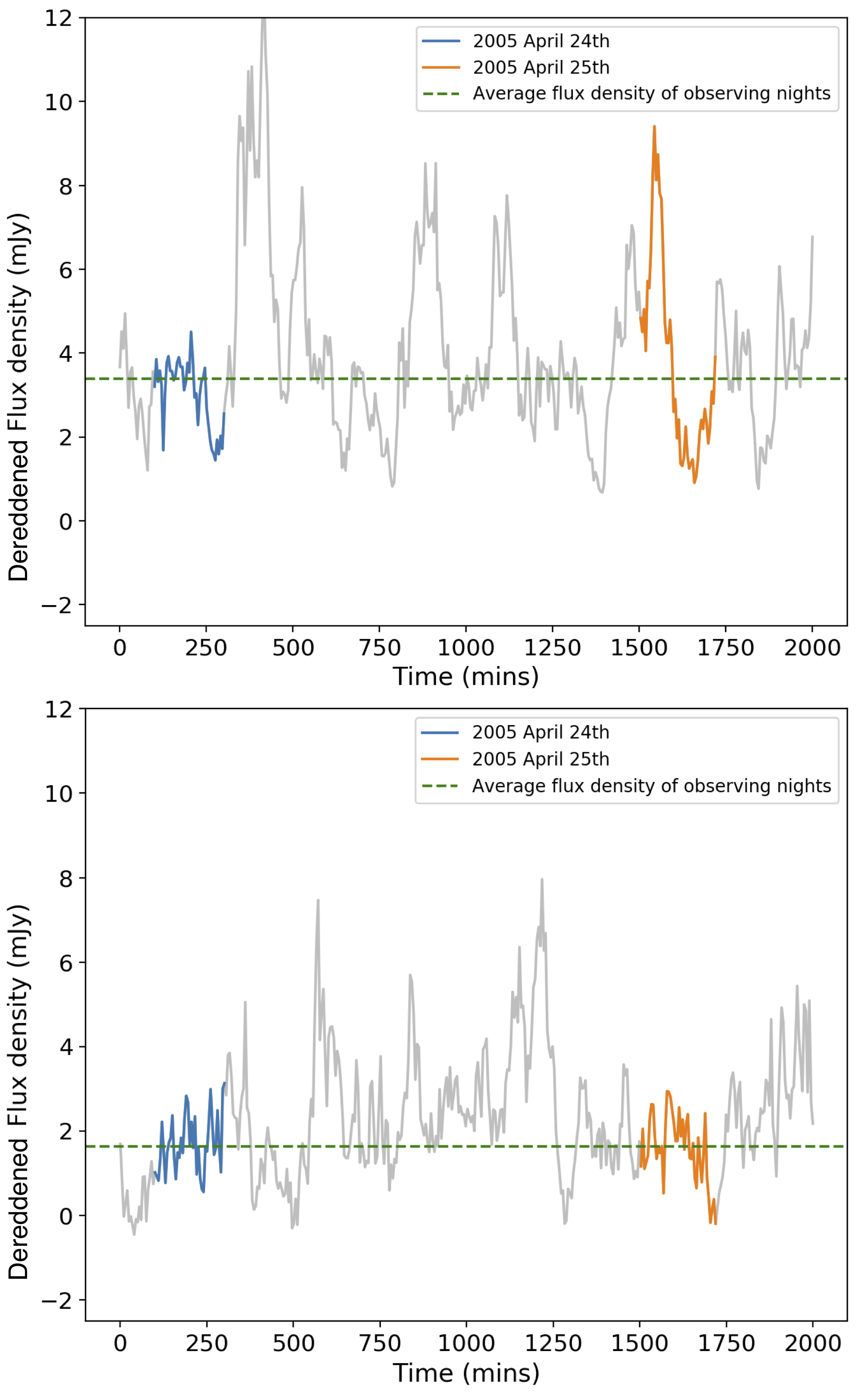}
\caption{Two examples of the simulated Sgr A* light curve (grey curve) over the two real observing nights in April 2005 (blue and orange parts respectively). The simulated light curves were generated following the modeled power spectrum density and a log-normal flux distribution, and using the same time series sampling as the real observations (see section \ref{sec:strucf}). The horizontal line shows the average flux density from the two observing nights (blue and orange), which imitates the final combined image.}
\label{fig:lc_sim_ex}
\end{figure}

\begin{figure*}
\centering
\includegraphics[width=7.2in]{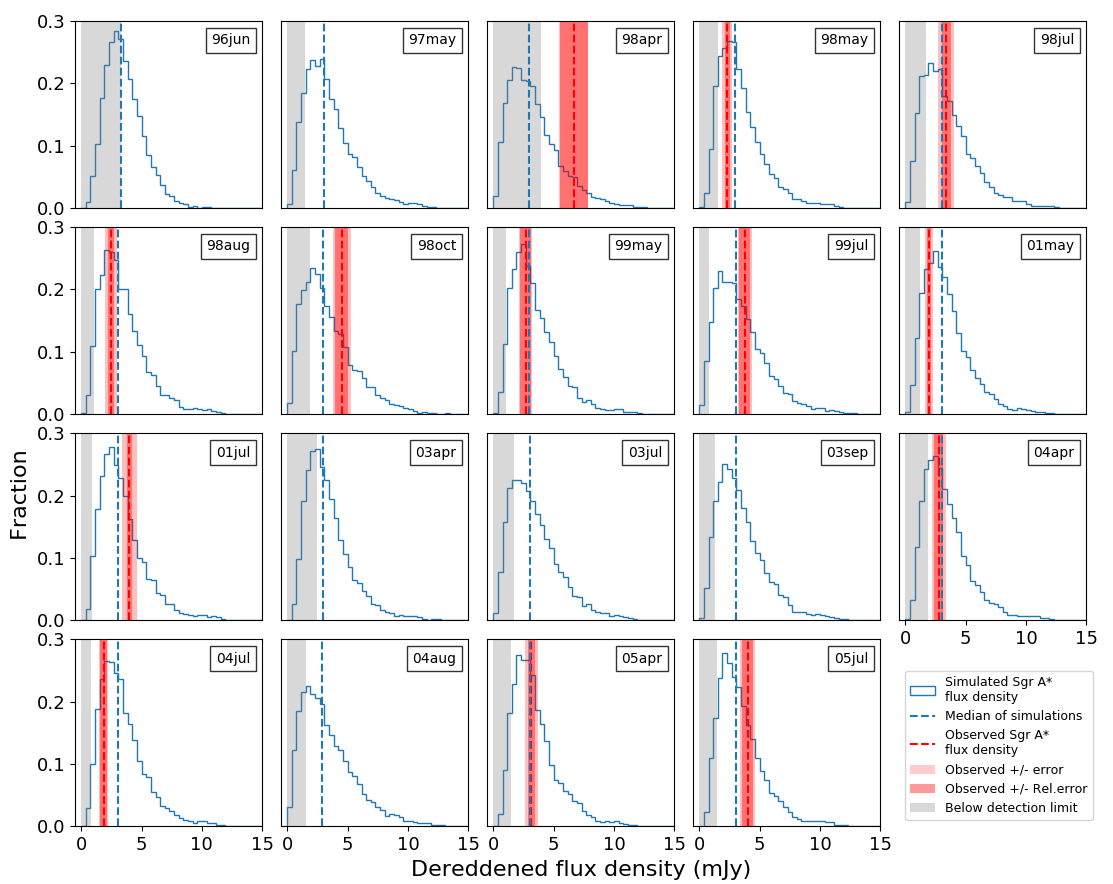}
\caption{The distribution of average flux densities of 10,000 simulated light curves (section \ref{sec:simu_method}) compared to the observations of Sgr A*. We probed 19 available epochs with either a detection or a detection limit. Blue curve presents the histogram of 10,000 average flux densities with median shown in blue dashed line. Red dashed line with bands marks the observed Sgr A* dereddened flux density with errors (light: total photometric error; dark: relative photometric error only, see Appendix \ref{sec:photometry}). The grey shaded region in each epoch presents the fluxes lower than the detection limit (defined in Appendix \ref{sec:lowerthreshold}).}
\label{fig:lc_hist}
\end{figure*}

\begin{figure}
\centering
\includegraphics[width=3.4in]{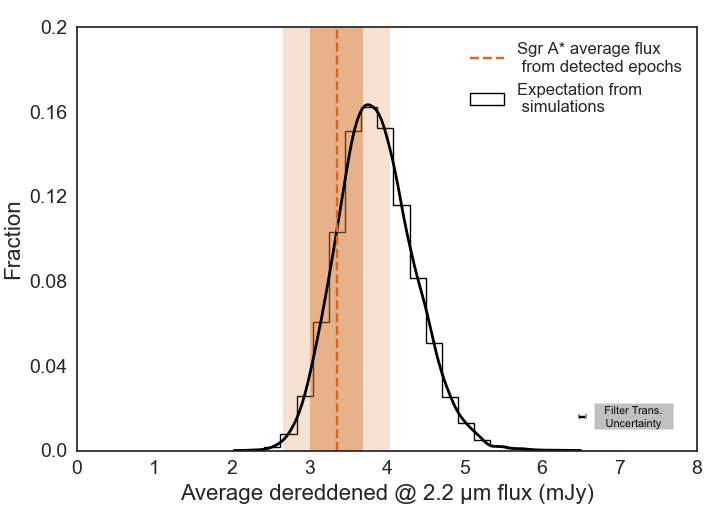}
\caption{Sgr A* average flux from early observations (1996 - 2005) and its expectation from simulations based on more recent observations (2006 - 2017). The orange line with band marks the average dereddened flux calculated from 13 detected epochs with errors (dark: $\sigma$$/$$\sqrt{N_{detections}}$; light: $2\sigma$$/$$\sqrt{N_{detections}}$). The histogram with corresponding kernel density estimation presents the distribution of expected average flux density from simulations based on the model in \citet{Witzel et al. 2018}. The error bar on the right bottom shows the filter transformation uncertainty of 0.04 mJy. Sgr A* has had similar brightness over two decades.}
\label{fig:flux_ave}
\end{figure}

\begin{figure}
\centering
\includegraphics[width=3.4in]{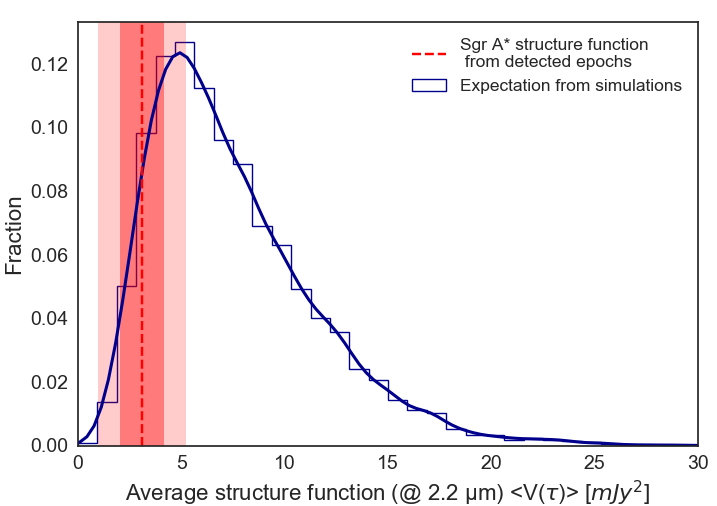}
\caption{Sgr A* average structure function $V(\tau)$ from observations and its expectation from simulations. The red line with band marks the average structure function calculated from 13 detected epochs with 1 $\sigma$ error and 2 $\sigma$ error respectively (see red points in Figure \ref{fig:sf}). The blue histogram with corresponding kernel density estimation presents the distribution of expected average structure function from simulations based on the model in \citet{Witzel et al. 2018} (see red bands in Figure \ref{fig:sf}). Sgr A* has had similar variability characteristics over two decades. }
\label{fig:struct_ave}
\end{figure}

\subsection{Long timescale variability of Sgr A*-IR} \label{sec:strucf}

While the short-term variability of Sgr A* in the near-infrared (NIR) is well characterized as a red-noise process, the long timescale variation has not been well probed. In order to explore if the observed long timescale variability shown in this work is consistent with models derived from shorter timescales, we simulate the NIR Sgr A* light curves with model presented in \citet{Witzel et al. 2018} (see section \ref{sec:simu_method}). Section \ref{sec:simtoobs} presents the comparison of the simulations to the observations. See section \ref{sec:breaktime} for further discussion of the characteristic break timescale.

\subsubsection{Light curve simulations}\label{sec:simu_method}

This model contains two key components. The first component describes the temporal characteristics using the power spectral density (PSD), which is modeled as a broken power law. 

\begin{equation}\label{equ:psdp}
    PSD(f) \propto \left \{
                     \begin{tabular}{ccc}
                     $f^{-\gamma_0}$ & for &  f < $f_b$ \\
                     $f^{-\gamma_1}$ & for &  f $\geq$ $f_b$ 
                     \end{tabular}
                   \right .
\end{equation}

where $\gamma_0 = 0$ (assumed), $\gamma_1 = 2.1 \pm 0.1$, $f_b = (4.1 \pm 0.7) * 10^{-3}$ $min^{-1}$ (which corresponds to a timescale of $\tau \sim 245 $ minutes). See Figure \ref{fig:psd_model} for the modeled PSD.

The second component describes the distribution of fluxes with a log-normal probability density function (PDF).

\begin{equation}\label{equ:pdfl}
\begin{split}
    \mathcal{P}[F|(\mu_{logn}, \sigma_{logn})] = \\
    (\sqrt{2\pi}F\sigma_{logn})^{-1} \cdot exp(- \frac{[ln(\frac{F(K)}{mJy}) - \mu_{logn}]^2}{\sqrt{2}\sigma_{logn}^2})
\end{split}
\end{equation}
where Sgr A* flux density $F \in [0,\infty]$, log-normal mean in K-band $\mu_{logn} \in [-\infty, +\infty]$, and log-normal standard deviation in K-band $\sigma_{logn} \in [0, \infty]$.  

Following the method in \citet{Timmer Koenig 1995} and using the modeled parameters in \citet{Witzel et al. 2018}, we are able to generate the simulated light curves of Sgr A* using the same time series sampling as the real observations following the time of data cubes. Here we assigned the time of each data cube to be the time of first frame in that cube. We have tested the effects of only using the cube time series instead of frame time series if considering time delay between each frame. No significant differences were found. We therefore use the cube sampling for computaional efficiency. To create each simulated light curve (with observed flux density), we added Gaussian-distributed noise ($\sigma = 0.035$ mJy, average uncertainty from speckle observations, see section \ref{sec:detections13}). In order to convert from observed fluxes at Kp (NIRC2 AO instrument used in this model from \citet{Witzel et al. 2018}) to our observations at K (NIRC Speckle instrument), we did filter transformation for Sgr A* of NIRC2 Kp - NIRC K = 0.367 mag$^{+0.01}_{-0.02}$ (see Table \ref{tab:filtersgra}), similar to the bandpass correction described in section \ref{sec:dered} and Appendix \ref{sec:calib}. As a final step, the dereddended flux of a simulated light curve was obtained following the process presented in section \ref{sec:dered}. See Figure \ref{fig:lc_sim_ex} for the examples of the final simulated light curve with dereddened flux.

\subsubsection{Comparison of the simulations to the observations}\label{sec:simtoobs}
We extracted the average flux density of every single simulated light curve (see section \ref{sec:simu_method}), in order to imitate each real observational image which combines and averages the fluxes from all data cubes. For each epoch, we repeated the simulation 10,000 times using the posterior values from \citet{Witzel et al. 2018}. See Figure \ref{fig:lc_hist} for the distribution of the average dereddened flux density of simulated light curves in 19 available epochs\footnote{with either a detection or a detection limit}, and Sgr A* observations for comparison.

Then we used simulated Sgr A* light curves to calculate the expectation of Sgr A* average flux and average structure function of our available epochs based on the model described above. In order to take into account the effect of detection limit of observations in simulations, we calculate the expectation of Sgr A* average flux and average structure function only with the simulated fluxes higher than the detection limit in that epoch. The steps are as following:
\begin{enumerate}[noitemsep,nolistsep]
  \item Probe 19 available epochs which have either a detection or a detection limit. 
  \item Among all 10,000 simulations, for each one set of simulated flux densities from all 19 epochs, mark the epoch if the simulated flux density passes the corresponding Sgr A* detection limit. Calculate one average flux density and one series of structure functions (for all possible pairs of time lags) with only marked epochs. Then generate one Sgr A* average variance. 
  \item Repeat step (2) for all 10,000 sets of simulations. The number of epochs used to generate each average flux density and series of structure functions varies depending on how many simulated flux densities pass the detection limit in that round. 
\end{enumerate}

Figure \ref{fig:flux_ave} presents the comparison of Sgr A* average dereddened flux from 13 detected epochs and its expectation calculated above from simulations based on the model in \citet{Witzel et al. 2018}. Figure \ref{fig:struct_ave} presents average structure function of Sgr A* detections and its expectation from simulations. The observed Sgr A* average flux density and average structure function are consistent with the predictions which are modeled from \citet{Witzel et al. 2018} with a power-law power-spectral density (PSD) and log-normal flux distribution. These results show that Sgr A* long-term variability status in the past (1998 - 2005) is well consistent with the extrapolation from shorter timescale AO-based observations at later time. Sgr A* has had similar brightness and variability characteristics over two decades.

\subsubsection{Characteristic break timescale}\label{sec:breaktime}

In this work, the flat structure function of Sgr A* calculated from observations indicates that there is no need for a second PSD break in the longer timescale we investigated here. Any significant increase of power in the PSD between $\sim$ 80 days and 7 years can be excluded. Previous studies of timescales from minutes to hours presented a break timescale in the NIR PSD of Sgr A* (\citealt{Meyer et al. 2008,Meyer et al. 2009,Do et al. 2009,Eckart et al. 2006b}) and the latest analysis \citep{Witzel et al. 2018} reports a correlation timescale of $\sim 245$ minutes. Our result is consistent with the assumption of a zero-slope PSD after the correlation timescale based on the model from \citet{Witzel et al. 2018}. Therefore, the 245 minute timescale remains the only confirmed break timescale in the NIR PSD of Sgr A*. Beyond this break timescale, Sgr A* appears to be uncorrelated of time, and the amplitude of the variations stop to increase.  

The speckle era of Keck (1995 - 2005) we explored in this work overlaps with some AO datasets between 2002 and 2005 (VLT NAOS/CONICA 2002 - 2005; Keck NIRC2 2004 - 2005). The results of overlapping AO datasets do not show any significant deviations from the average flux obtained with speckle datasets in the same period. \citet{Witzel et al. 2018} has summarized and reported the analysis based on AO datasets (VLT NAOS/CONICA 2003 - 2010; Keck NIRC2 2004 - 2016). The results obtained from AO measurements appear to be consistent with the speckle results reported in this work.

The characteristic timescale of the X-ray variability of active galactic nuclei (AGNs) and BH X-ray binaries (BHXRBs) has been similarly investigated (see, e.g., \citealt{Uttley et al. 2002,Markowitz et al. 2003,Uttley}). Previous studies hypothesized that the characteristic break timescales (the break frequency; observed in BHXRBs) of AGN scale linearly with the mass of BH with a correction factor of bolometric luminosity of the accretion flow \citep{McHardy et al. 2006}. And this led to the conclusion that AGNs are scaled-up galactic BHs. Therefore, the variability of the SMBH at the galactic center, Sgr A*, serves well as the most under-luminous BH system to test the scaling relationship of AGNs (\citealt{Meyer et al. 2009,Witzel et al. 2018}).

\section{Conclusions}\label{sec:con}

The long-term variability of the supermassive black hole at the Galactic Center, Sgr A*, has been studied with the analysis of speckle data (1995-2005) obtained from the W. M. Keck 10 m telescope. The application of the speckle holography technique enables us to investigate Sgr A* with deeper detections than in any previous work. This study presents the first near-infrared (NIR) detection of Sgr A* prior to 2002. We are able to monitor the long-term variability of Sgr A* in the NIR with a time baseline of 7 years with analysis for astrometry and photometry of Sgr A*. We present a Sgr A* light curve from 1998 to 2005 which indicates Sgr A* was stable and showed no extraordinary flux excursions during this time. The average observed magnitude of Sgr A* as obtained from the last seven years (1998 - 2005) of speckle holography datasets is $K = 16.0$ $\pm$ $0.4$ with average relative photometric uncertainty of 0.1 mag, corresponding to the average observed flux density of $0.35$ $\pm$ $0.13$ mJy with average uncertainty of $0.04$ mJy. The average dereddened flux density is $3.4$ $\pm$ $1.2$ mJy with total average photometric uncertainty of $0.6$ mJy. The results agree very well with the average observed AO measurements of $K = 16.1$ $\pm$ $0.3$ (2005 - 2017), and are consistent with the extrapolation modeled from AO-based shorter timescale studies. Sgr A* is quite stable without significant change in this time baseline of 7 years based on the structure function timing analysis, which indicates that 245 minutes still remains the dominant break timescale. Based on the results, the periapse passage of the object G1 did not result in any measurable change of the mean accretion rate onto Sgr A*.

\section{Acknowledgement} \label{sec:ack}
The primary support for this work was provided by NSF, through grant AST-1412615, and UCLA, through faculty salaries. Additional support was received from the Heising-Simons Foundation, the Levine-Leichtman Family Foundation, the Preston Family Graduate Fellowship (held by B. N. S and A. G.), UCLA Galactic Center Star Society, UCLA $Cross-disciplinary$ $Scholars$ $in$ $Science$ $and$ $Technology$ (CSST) Fellowship, NSF $Research$ $Experiences$ $for$ $Undergraduates$ (REU) Grant No. PHY-1460055, the European Union's Seventh Framework Programme (FP7/2007-2013) / ERC grant agreement n° [614922](R.S.) and financial support from the State Agency for Research of the Spanish MCIU through the "Center of Excellence Severo Ochoa" award for the Instituto de Astrofísica de Andalucía (SEV-2017-0709)(R.S.). This research was based on data products from the $Galactic$ $Center$ $Orbit$ $Initiative$ (GCOI), which is hosted at UCLA and which is a key science program of the Galactic Center Collaboration (GCC). These data products were derived from data originally obtained from W. M. Keck Observatory. The W. M. Keck Observatory is operated as a scientific partnership among the California Institute of Technology, the University of California, and the National Aeronautics and Space Administration. The authors wish to recognize that the summit of Mauna Kea has always held a very significant cultural role for the indigenous Hawaiian community. We are most fortunate to have the opportunity to observe from this mountain. The Observatory was made possible by the generous financial support of the W. M. Keck Foundation.

\begin{deluxetable*}{lrccccccccc}[t]
\tabletypesize{\footnotesize}
\tablecolumns{9}
\tablewidth{0pt}
\tablecaption{Summary of Speckle Holography Observations$^{a}$
\label{tab:speckle1}}

\tablehead{
\multicolumn{2}{c}{Date} & \colhead{Frames} & \colhead{$K_{lim}$ $^{b}$} & \colhead{$N_{Real}$} & \colhead{$N_{pix}$ $^{c}$} &  \multicolumn{2}{c}{Sgr A* Pos. with respect to FoV  $^{e}$}
 & \colhead{Original} & \colhead{K Systematic Phot. } & \colhead{K Relative Phot.} \\ \cline{1-2} \cline{7-8} \colhead{(U.T.)} & \colhead{(Decimal)} & \colhead{Used} & \colhead{(mag)} & \colhead{} & \colhead{} & \colhead{$\Delta$R.A. (arcsec)} & \colhead{$\Delta$Dec. (arcsec)}
 & \colhead{Refs.$^{f}$} & \colhead{Zero-point Error$^{g}$ (mag)} & \colhead{Error$^{h}$ (mag)}}
\startdata
1995 Jun 9-12 & 1995.439 & 5265 & 17.0 & 41 & 108042 & -0.52 & 0.01 & 1 & 0.24 & 0.04\\
1996 Jun 26-27 & 1996.485 & 2283 & 15.8 & 49 & 82505 & -1.22 & -0.29 & 1 & 0.14 & 0.09\\
1997 May 14 & 1997.367 & 3426 & 16.8 & 51 & 92467 & -0.90 & -0.15 & 1 & 0.09 & 0.03\\
1998 Apr 2-3 & 1998.251 & 1718 & 15.8 & 39 & 95816   & -0.57 & -0.15 & 2 & 0.07 & 0.06\\
1998 May 14-15 & 1998.366 & 7675 & 16.8 & 45 & 102328  & -0.45 & -0.10 & 2 & 0.17 & 0.04\\
1998 Jul 3-5 & 1998.505 & 2040 & 16.4 & 43 & 116557  & 0.04 & 0.00 & 2 & 0.18 & 0.05\\
1998 Aug 4-6 & 1998.590 & 11032 & 17.1 & 47 & 109269  & -0.41 & 0.04 & 2 & 0.18 & 0.04\\
1998 Oct 9,11 & 1998.771 & 2000 & 16.6 & 45 & 97215 & 0.80 & 0.05 & 2 & 0.12 & 0.03\\
1999 May 2-4 & 1999.333 & 9423 & 17.2 & 52 & 107882 & -0.45 & -0.21 & 2 & 0.12 & 0.06\\
1999 Jul 24-25 & 1999.559 & 5690 & 17.4 & 54 & 100567  & -0.46 & -0.09 & 2 & 0.11 & 0.04\\
2000 Apr 21 & 2000.305 & 651 & 15.7 & 56 & 96248  & 0.84  & 0.11 & 3 & 0.09 & 0.04\\
2000 May 19-20 & 2000.381 & 15581 & 17.5 & 55 & 96853    & -0.74 & -0.24 & 3 & 0.08 & 0.03\\
2000 Jul 19-20 & 2000.584 & 10668 & 17.0 & 63 & 86452   & -0.93 & -0.12 & 3 & 0.15 & 0.04\\
2000 Oct 18 & 2000.797 & 2215 & 16.2 & 52 & 82315 & -0.80 & -0.42 & 3 & 0.09 & 0.05\\
2001 May 7-9 & 2001.351 & 6662 & 17.2 & 64 & 85028   & -0.46 & -0.20 & 3 & 0.17 & 0.02\\
2001 Jul 28-29 & 2001.572 & 6634 & 17.4 & 74 & 96872  & -0.15 & -0.22 & 3 & 0.15 & 0.02\\
2002 Apr 23-24 & 2002.309 & 13440 & 17.5 & 74 & 96953 & -0.59 & -0.17 & 3 & 0.12 & 0.05\\
2002 May 23-24 & 2002.391 & 11834 & 17.6 & 72 & 98552   & -0.85 & -0.08 & 3 & 0.14 & 0.05\\
2002 Jul 19-20 & 2002.547 & 4139 & 16.8 & 69 & 99994   & -0.63 & -0.39 & 3 & 0.17 & 0.05\\
2003 Apr 21-22 & 2003.303 & 3644 & 16.4 & 58 & 90963  & -0.32 & -0.40 & 3 & 0.18 & 0.06\\
2003 Jul 22-23 & 2003.554 & 2894 & 16.8 & 65 & 87265   & -0.54 & -0.24 & 3 & 0.08 & 0.01\\
2003 Sep 7-8 & 2003.682& 6296 & 17.1 & 74 & 95367  & -0.53 & -0.44 & 3 & 0.14 & 0.03\\
2004 Apr 29-30 & 2004.327 & 6169 & 16.8 & 58 & 125423 & -0.71 & -0.21 & 4 & 0.17 & 0.04\\
2004 Jul 25-26 & 2004.564 & 13071 & 17.4 & 80 & 99819  & -0.61 & -0.41 & 4 & 0.15 & 0.04\\
2004 Aug 29 & 2004.660 & 2284 & 16.8 & 63 & 96172  & -0.09 & 0.66 & 4 & 0.14 & 0.02\\
2005 Apr 24-25 & 2005.312 & 9553 & 17.1 & 70 & 105715 & -0.36 & -0.16 & 5 & 0.14 & 0.05\\
2005 Jul 26-27 & 2005.566 & 5606 & 16.8 & 84 & 108360 & -0.26 & -0.41 & 5 & 0.12 & 0.04\\
\enddata
\tablecomments{\\
$^{a}$ All numbers given in the table are based on speckle holography version 2\_2 (see section \ref{sec:dataAna}). \\
$^{b}$ $K_{lim}$ is the magnitude that corresponds to the 95$^{th}$ percentile of all $K$ magnitudes in the sample of real stars in the central $2'' \times 2''$ region (see section \ref{sec:lowerthreshold}). \\
$^{c}$ $N_{pix}$ refers to the number of pixels in a given image that meet a .8 of maximum frames used criteria. \\
$^{d}$ $N_{ref}$ refers to the number of reference stars used to align the epoch of data.  \\
$^{e}$ The center of the Field of View (FoV) is the weighted average of the detector pixels, with the weight being the number of frames used squared at each pixel in the final image. The offsets reported above are ($\vec{x}_{Sgr A*} - \vec{x}_{center FoV}$).\\
$^{f}$ 1: \citep{Ghez et al. 1998}; 2: \citep{Ghez et al. 2000}; 3: \citep{Ghez et al. 2005a}; 4: \citep{Lu et al. 2005}; 5: \citep{Rafelski et al. 2007} \\
$^{g}$ Systematic photometric zero-point errors were calculated after performing initial photometric system calibration described in Appendix \ref{sec:calib}. The average zero-point uncertainty $\overline{\sigma}_{zp}$ is 0.14 mag in NIRC K bandpass.\\
$^{h}$ Relative photometric zero-point errors were determined by the relative photometry calibration using the stable calibrators (see Appendix \ref{sec:calibrela}). The average zero-point uncertainty $\overline{\sigma}_{zp}$ for the relative photometry is 0.04 mag in NIRC K bandpass.\\
}
\end{deluxetable*}

\begin{longrotatetable}
\tablenum{2}
\begin{deluxetable*}{L{2cm}C{1.6cm}C{1.6cm}C{1.6cm}C{1.6cm}C{1.6cm}C{1.6cm}C{1.6cm}C{1.6cm}C{1.6cm}C{1.6cm}C{0.1cm}}
\tablecaption{Summary of Sgr A* Measurements\label{table:chartable}}
\tablewidth{700pt}
\tabletypesize{\scriptsize}
\tablehead{
    Date(UT) & Date (Epoch) & Bootstrap Fraction Cut$^{a}$ & Source Detection Limit$^{b}$ (mag) & Source Confusion with Sgr A* & K (mag) & K Error (mag) & Dereddened Flux (mJy) & Dereddened Flux Error (mJy) &  Fraction of Bootstraps Detected & $\Delta$T$^{c}$ (mins)\\
}
\startdata
1995 June 9-12    & 1995.439 &0.26 &17.0 & S0-19   &\nodata &\nodata &\nodata& \nodata &\nodata &\nodata & \\
1996 June 26-27   & 1996.485 &0.24 &16.0 & \nodata &\nodata &\nodata &\nodata& \nodata &\nodata &\nodata & \\
1997 May 14       & 1997.367 &0.22 &16.9 & \nodata &\nodata &\nodata &\nodata& \nodata &\nodata &\nodata & \\
1998 April 2-3    & 1998.251 &0.14 &15.8 & \nodata & 15.2   &0.2     & 6.7   & 1.2     &0.64    & 352 &\\
1998 May 14-15    & 1998.366 &0.02 &16.8 & \nodata & 16.4   &0.1     & 2.3   & 0.4     &0.48    & 1653 &\\
1998 July 3-5     & 1998.505 &0.12 &16.7 & \nodata & 16.0   &0.1     & 3.4   & 0.7    &0.34    & 1643 &\\
1998 August 4-6   & 1998.59  &0.02 &17.2 & \nodata & 16.3   &0.1     & 2.5   & 0.5    &0.66    & 1624 &\\
1998 October 9,11    & 1998.771 &0.06 &16.6 & \nodata & 15.7   &0.1     & 4.5   & 0.7     &0.44    & 2952 &\\
1999 May 2-4      & 1999.333 &0.06 &17.2 & \nodata & 16.2   &0.2     & 2.7   & 0.5     &0.82    & 2949 &\\
1999 July 24-25   & 1999.559 &0.02 &17.5 & \nodata & 15.9   &0.1     & 3.8   & 0.6     &0.88    & 1527 &\\
2000 April 21     & 2000.305 &0.14 &15.7 & S0-16   &\nodata &\nodata &\nodata& \nodata &\nodata &\nodata & \\
2000 May 19-20    & 2000.381 &0.02 &17.6 & S0-16   &\nodata &\nodata &\nodata& \nodata &\nodata &\nodata & \\
2000 July 19-20   & 2000.548 &0.14 &17.3 & S0-16   &\nodata &\nodata &\nodata& \nodata &\nodata &\nodata & \\
2000 October 18   & 2000.797 &0.12 &16.4 & S0-16   &\nodata &\nodata &\nodata& \nodata &\nodata &\nodata & \\
2001 May 7-9      & 2001.351 &0.08 &17.1 & \nodata & 16.6   &0.1     & 2.0   & 0.3     &0.81    & 1443 &\\
2001 July 28-29   & 2001.572 &0.02 &17.4 & \nodata & 15.8   &0.1     & 4.0   & 0.6     &0.96    & 1702 &\\
2002 April 23-24  & 2002.309 &0.06 &17.4 & S0-2    &\nodata &\nodata &\nodata& \nodata &\nodata &\nodata & \\
2002 May 23-24    & 2002.391 &0.02 &17.8 & S0-2    &\nodata &\nodata &\nodata& \nodata &\nodata &\nodata & \\
2002 July 19-20   & 2002.547 &0.22 &16.9 & S0-2    &\nodata &\nodata &\nodata& \nodata &\nodata &\nodata & \\
2003 April 21-22  & 2003.303 &0.04 &16.3 & S0-38   &\nodata &\nodata &\nodata& \nodata &\nodata &\nodata & \\
2003 July 22-23   & 2003.554 &0.08 &16.7 & \nodata &\nodata &\nodata &\nodata& \nodata &\nodata &\nodata & \\
2003 September 7-8& 2003.682 &0.18 &17.0 & \nodata &\nodata &\nodata &\nodata& \nodata &\nodata &\nodata & \\
2004 April 29-30  & 2004.327 &0.02 &16.6 & \nodata & 16.2   &0.1     & 2.8   & 0.6      &0.21    & 1578 &\\
2004 July 25-26   & 2004.564 &0.04 &17.5 & \nodata & 16.7   &0.2     & 1.9   & 0.4     &0.75    & 1572 &\\
2004 August 29    & 2004.66  &0.06 &16.8 & \nodata &\nodata &\nodata &\nodata& \nodata &\nodata & \nodata &\\
2005 April 24-25  & 2005.312 &0.06 &16.9 & \nodata & 16.0   &0.1     & 3.2   & 0.5     &0.95    & 1633 &\\
2005 July 26-27   & 2005.566 &0.16 &16.9 & \nodata & 15.8   &0.1     & 4.0   & 0.6     &0.66    & 1654 &\\
\enddata
\tablecomments{\\
$^{a}$ Bootstrap fraction cut of a real source detection in each epoch is obtained at which the probability of a false detection within within 10 mas radius is always $< 1\%$.\\
$^{b}$ Sgr A* detection limit is the 95$^{th}$ percentile of all $K$ magnitudes in the sample which includes all sources with a bootstrap fraction higher than the Bootstrap Fraction Cut in the central $2'' \times 2''$ field of view.\\
$^{c}$ Time duration of the observations that contributes to the final speckle image for Sgr A* detections. 
}

\end{deluxetable*}
\label{table:SpeckleHolo}
\end{longrotatetable}

\appendix
\begin{figure}
    \centering
    \includegraphics[width=7.2in]{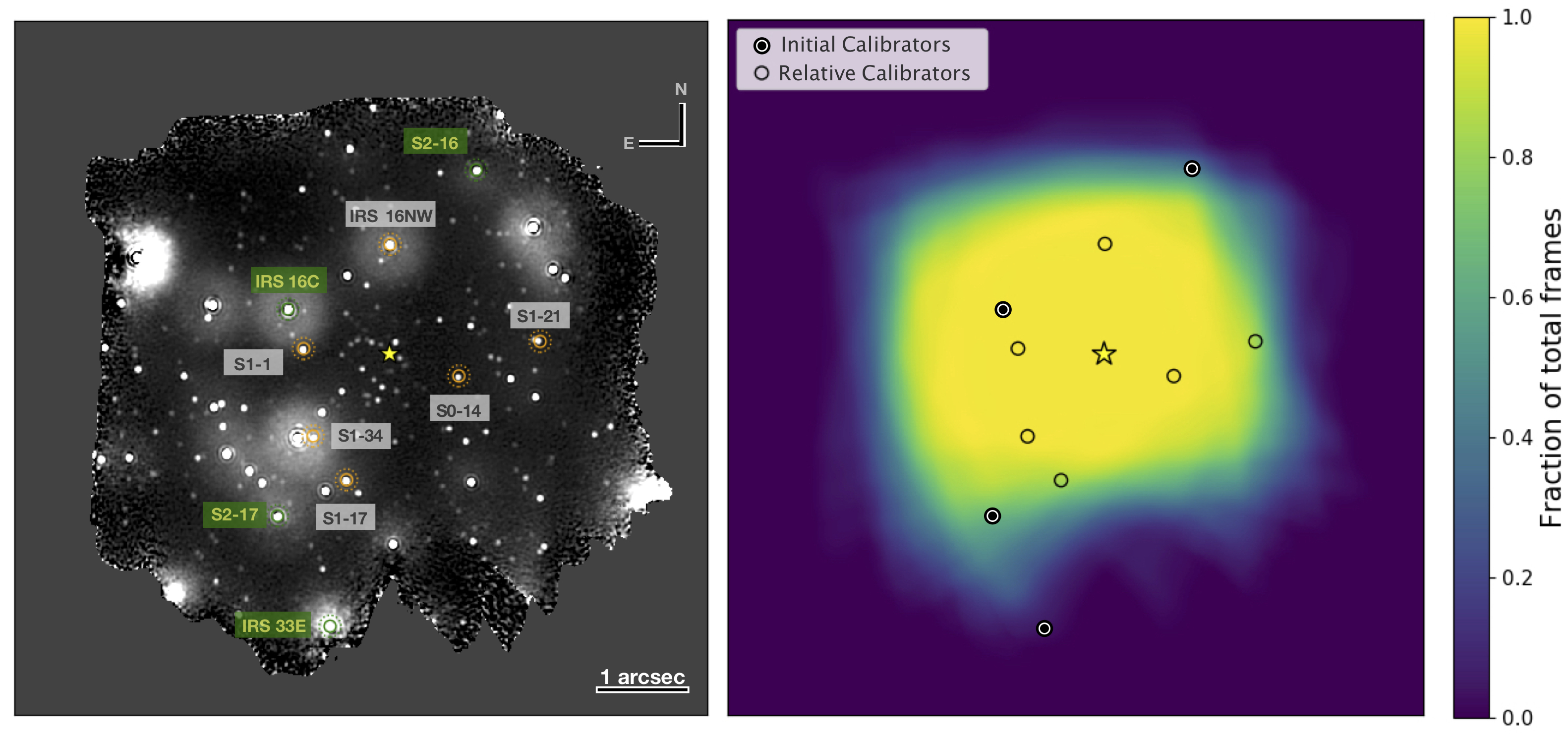}
    \caption{Location of photometric calibrators. Left: Background is the speckle holography image from April 2002 observation. The 4 initial photometric calibrators used in photometric system calibration are marked with green circles. The 6 relative photometric calibrators are marked with orange circles. The yellow star symbol shows Sgr A*'s position. Right: The fraction of total frames used in each pixel for the final image from April 2002 observation. Filled circles show initial photometric calibrators, and the open circles show relative photometric calibrators. The relative calibrators are chosen to be isolated stars that uniformly cover the field of view with minimal edge effects.}
    \label{fig:calibrators}
\end{figure}

\section{Photometry used in this work}\label{sec:photometry}

In order to obtain photometry for sources from the speckle holography images, we performed a two-step procedure. In the first step ($\S$ \ref{sec:calib}), the photometric systematic scale is established with an uncertainty of 0.14 mag (1$\sigma$). In the second step ($\S$ \ref{sec:calibrela}), we select a set of stable secondary photometric calibrators to improve the relative photometry to $\pm 0.04$ mag (1$\sigma$).

\begin{figure}
    \centering
    \includegraphics[width=3.6in]{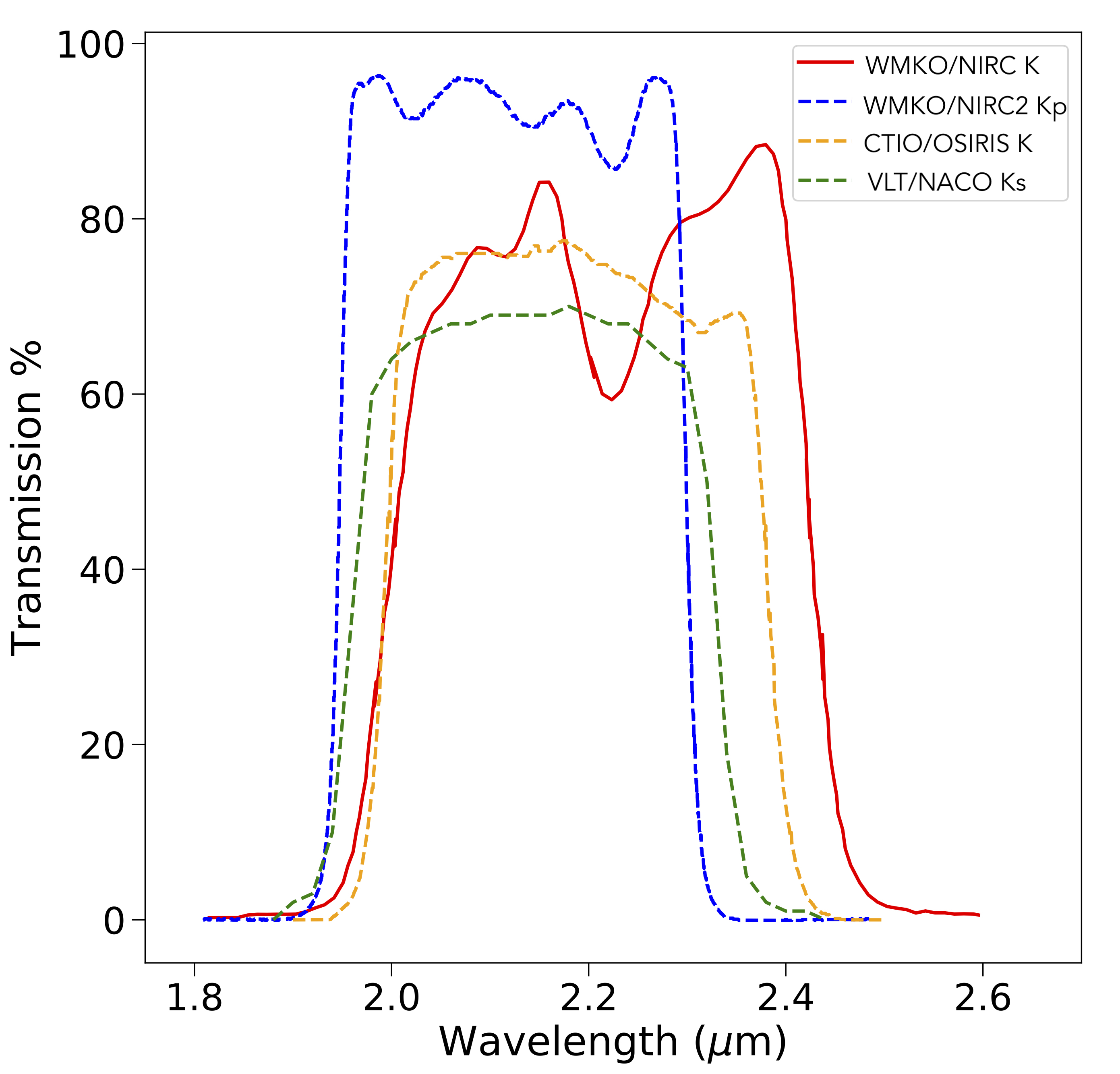}
    \caption{Comparison of the 2 $\mu$m bandpass filter used in this work. The transmission curves for the WMKO/NIRC/K (this work), WMKO/NIRC2/Kp (\citealt{Gautam et al. 2018}), CTIO/OSIRIS/K (\citealt{Blum et al. 1996}), and VLT/NAKO/Ks (\citealt{Schodel et al. 2010}). Owing to the different transmissions, there are photometric offsets between the filters (see Appendix \ref{sec:calib}).}
    \label{fig:bandpass}
\end{figure}

\subsection{Photometric system Calibration}\label{sec:calib}

We perform photometric system calibration using four initial calibration stars (IRS 16C, IRS 33E, S2-16 and S2-17). These stars are the only ones that have both reference flux measurements as reported by \citet{Blum et al. 1996} and are located within our field of view. As Figure \ref{fig:calibrators} shows, IRS 16C is ideally located close to the center of the field of view and therefore measured in every epoch. S2-17 and S2-16 are measured in almost every epoch, and in each epoch typically have more than half of the frames obtained. In contrast, IRS 33E is much closer to the edge of the final field of view and detected in only $2/3$ of the epochs, and in these epochs typically has $1/3$ of the frames. 

There are three considerations made to convert Blum's measurements into flux predictions for the speckle holography measurements made with NIRC. First and most importantly, we applied an aperture correction to the Blum's measurements to account for the low resolution of their measurements. With $\sim 1''$ seeing, Blum's measurements include neighboring stars that are resolved in our speckle holography observations. Therefore, we did aperture correction by subtracting the fluxes of nearby sources within the radius of $\sim 0.5''$ aperture. The correction ranges from 0 to 0.3 magnitudes. Second, owing to the slight differences between the bandpass used in \citet{Blum et al. 1996} and our NIRC K instrument (see Figure \ref{fig:bandpass}), there are photometric offsets between the filters. These offsets were calculated by convolving the extincted stellar model atmospheres with the filter functions (see Appendix A in \citet{Gautam et al. 2018} for more details). The offsets range from 0.06 to 0.11 magnitudes. Third, the uncertainties in the predicted brightness of the four calibrators are increased by the known level of variability from the work in \citet{Gautam et al. 2018}. The additional uncertainties range from less than 0.03 to 0.07 magnitudes, which for each star is less than the uncertainty in the original Blum's measurements. Table \ref{tab:bandpass} summarizes all these considerations and provides the final predictions. 

The zeropoint for each epoch, $\text{zp}$, was calculated as a weighted mean of the ratios between the calibrators' measured instrumental flux and reference flux ($d_i = f_{i, \text{reference}} / f_{i, \text{measured}}$):
\begin{eqnarray}
    \text{zp} = \frac{\sum w_i * d_i}{\sum w_i}
\end{eqnarray}
Here, $w_i$ indicates the weight for calibrator $i$, derived from uncertainty in its reference flux ($\sigma_{f, i}$): $w_i = (\sigma_{f, i})^{-2}$.
Zero-point uncertainties in the photometric calibration for each observation epoch, $\sigma_{\text{zp}}$, were derived from the weighted standard deviation of flux differences (between the calibrator stars' reference magnitudes and measured magnitudes) then divided by $\sqrt{N_{Calibs.}- 1}$ (here $N_{Sys.Calibs.} = 4$).
\begin{eqnarray}
    \sigma_{\text{zp}} = \sqrt{\frac{\sum w_i (d_i - \text{zp})^2}{\sum w_i}} / \sqrt{N_{Calibs.}- 1}
\end{eqnarray}

\begin{figure}
    \centering
    \includegraphics[width=3.5in]{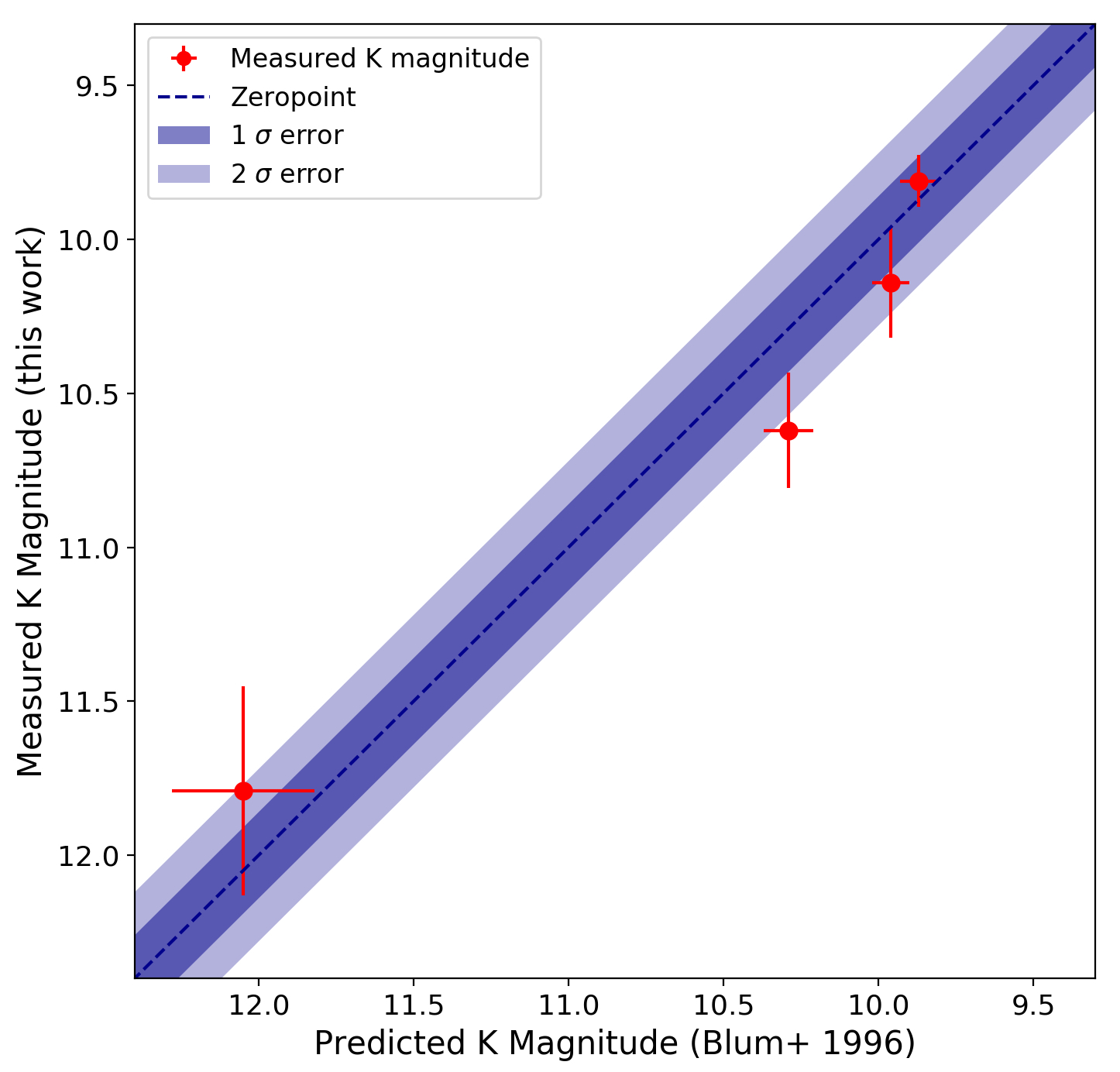}
    \caption{Comparison of our measured K magnitude (from photometric system calibration) to the predicted K magnitude (band-pass corrected from \citet{Blum et al. 1996}) for all 4 initial calibrators (Red points with errorbars). The resulting zeropoint uncertainty is 0.14 mag (1 $\sigma$). }
 
    \label{fig:zeropoint}
\end{figure}

Each epoch's zero-point uncertainty is reported in Table \ref{tab:speckle1}. Overall, we achieved an average zero-point uncertainty $\overline{\sigma}_{zp}$ for the photometric system calibration of 0.14 mag in NIRC K bandpass (see Figure \ref{fig:zeropoint}).

\begin{figure}
    \centering
    \includegraphics[width=4.2in]{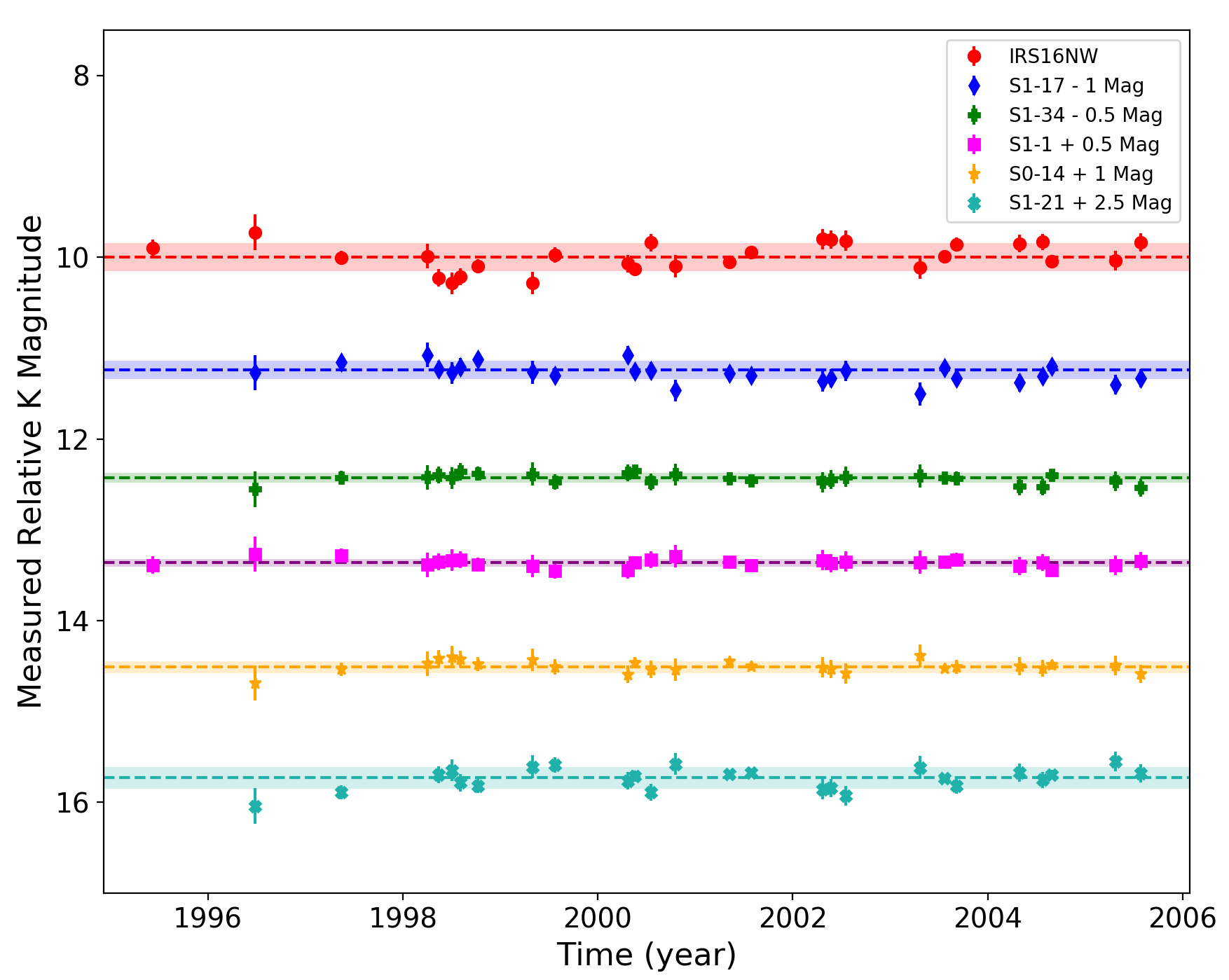}
    \caption{Light curve with measured relative K magnitude for each of the secondary calibrator star after relative photometric calibration. The line with band shows the weighted mean and the RMS across all detected epochs. The relative photometry has an average uncertainty of 0.04 mag (1 $\sigma$).}
    \label{fig:lc_sec}
\end{figure}

As a final step, we verify that our speckle holography measurements are at the same photometric system as \citet{Witzel et al. 2018} whose model is used to simulated the near-infrared Sgr A* light curves in section \ref{sec:strucf}. \citealt{Witzel et al. 2012} used and reported 13 stars as photometric calibrators of which 12 are contained in our studies (see Appendix \ref{sec:sourceanalyses}). These measurements were tied to the absolute Ks observations reported in \citet{Schodel et al. 2010}. Therefore, we transformed from the VLT NACO Ks to NIRC K (see Figure \ref{fig:bandpass} for different transmissions) with bandpass corrections similar to that process described above. Figure \ref{fig:k_ks} shows that our measured K magnitudes (after photometric system calibration) are highly consistent with the predicted K magnitudes (corrected from Ks photometric system), with an average difference of only 0.01 $\pm$ 0.09 mag (see Table \ref{tab:bandpass}).

\begin{table*}
    \tablenum{3}
    {
    \caption{Calibration Stars}\label{tab:bandpass}
    \begin{tabular}{cccrcccccc}
    \hline 
    \hline 
    Star & Calib. &N $^{a}$ &Varia.$^{b}$ & $K_{Blum+96}$& Aperture & $K_{NIRC}$-&$K_{NIRC, Pred.}^{d}$ & $K_{NIRC}$& $K_{NIRC}$   \\
    Name &Type &epoch &(mag) & &Corr.$^{c}$ ($\delta$mag) &$K_{Blum+96}$ & &  Init.$^{e}$ &Rel.$^{f}$  \\
    \hline
    
    IRS 16C & Phot. &27& 0.040  & 9.86  $\pm$ 0.05 & 0.07 $\pm$ 0.001 & -0.06 $\pm$ 0.01 &  9.87$\pm$0.06$^{g}$$\pm$0.04$^{h}$ & 9.81$\pm$0.04$\pm$0.14 & 9.79$\pm$0.08$^{i}$   \\
    
    IRS 33E & Phot. &18& < 0.035  & 10.02 $\pm$ 0.05 & 0 $\pm$ 0 & -0.06 $\pm$ 0.01 & 9.96$\pm$0.06$\pm$0.04  & 10.14$\pm$0.04$\pm$0.14  & 10.04 $\pm$0.07   \\
    S2-17   & Phot. &26&< 0.034  & 10.03 $\pm$ 0.07 & 0.32$\pm$ 0.007 & -0.06 $\pm$ 0.01 & 10.29$\pm$0.09$\pm$0.04  & 10.62$\pm$0.04$\pm$0.14 & 10.64$\pm$0.06   \\
    S2-16   & Phot. &25 &0.070   & 11.90 $\pm$ 0.22 & 0.26$\pm$0.001 & -0.11 $\pm$ 0.01 & 12.05$\pm$0.23$\pm$0.04 & 11.79$\pm$0.09$\pm$0.14  & 11.79$\pm$0.10   \\
    \hline
    IRS 16NW & Rel. &27 & < 0.031  & \nodata & \nodata & \nodata &\nodata & 10.00$\pm$0.04$\pm$0.14 & 10.00 $\pm$ 0.05   \\
    S1-17    & Rel. &26& < 0.030  & \nodata & \nodata & \nodata &\nodata & 12.24$\pm$0.04$\pm$0.14 & 12.24 $\pm$ 0.04  \\
    S1-34    & Rel. &26&< 0.029   & \nodata & \nodata & \nodata &\nodata & 12.92$\pm$0.04$\pm$0.14 &  12.93 $\pm$ 0.04 \\
    S1-1     & Rel.  &27&< 0.029  & \nodata & \nodata & \nodata &\nodata & 12.85$\pm$0.04$\pm$0.14 & 12.86 $\pm$ 0.04 \\
    S1-21    & Rel.  &25& < 0.030 & \nodata & \nodata & \nodata &\nodata & 13.26$\pm$0.05$\pm$0.14 &  13.23$\pm$ 0.05 \\
    S0-14    & Rel.\& Veri.  &26&< 0.028  & \nodata & \nodata & \nodata &13.50$\pm$0.03$\pm$0.07 & 13.50$\pm$0.04$\pm$0.14 &  13.51$\pm$ 0.05  \\
    \hline
    S0-13    & Veri.  &27& < 0.030  & \nodata & \nodata & \nodata &13.22$\pm$0.04$\pm$0.07 & 13.21$\pm$0.04$\pm$0.14 &  13.23$\pm$ 0.04 \\
    S0-6     & Veri.  &27& < 0.030 & \nodata & \nodata & \nodata &13.99$\pm$0.03$\pm$0.07 & 13.95$\pm$0.04$\pm$0.14 &  13.95$\pm$ 0.04 \\
    S0-12    & Veri.  &27& < 0.034  & \nodata & \nodata & \nodata &14.20$\pm$0.04$\pm$0.07 & 14.15$\pm$0.04$\pm$0.14 &  14.13$\pm$ 0.05 \\
    S0-4     & Veri.  &27& < 0.032  & \nodata & \nodata & \nodata &14.17$\pm$0.04$\pm$0.07 & 14.22$\pm$0.04$\pm$0.14 &  14.20$\pm$ 0.05 \\
    S1-10    & Veri.  &27& < 0.029  & \nodata & \nodata & \nodata &14.67$\pm$0.04$\pm$0.07 & 14.69$\pm$0.04$\pm$0.14 &  14.71$\pm$ 0.04 \\
    S0-31    & Veri.  &27& < 0.037  & \nodata & \nodata & \nodata &15.08$\pm$0.05$\pm$0.07 & 14.86$\pm$0.06$\pm$0.14 &  14.89$\pm$ 0.07 \\
    S1-33    & Veri.  &27& < 0.030  & \nodata & \nodata & \nodata &14.82$\pm$0.04$\pm$0.07 & 14.92$\pm$0.04$\pm$0.14 &  14.89$\pm$ 0.04 \\
    S0-11    & Veri.  &27& < 0.035  & \nodata & \nodata & \nodata &15.00$\pm$0.04$\pm$0.07 & 15.06$\pm$0.05$\pm$0.14 &  15.07$\pm$ 0.06 \\
    S1-6     & Veri.  &27& 0.069  & \nodata & \nodata & \nodata &15.25$\pm$0.08$\pm$0.07 & 15.35$\pm$0.08$\pm$0.14 &  15.44$\pm$ 0.08 \\
    S0-27    & Veri.  &26& 0.049  & \nodata & \nodata & \nodata &15.45$\pm$0.05$\pm$0.07 & 15.35$\pm$0.07$\pm$0.14 &  15.37$\pm$ 0.10 \\
    S1-31    & Veri.  &26 & < 0.036  & \nodata & \nodata & \nodata &15.50$\pm$0.05$\pm$0.07 & 15.51$\pm$0.05$\pm$0.14 &  15.51$\pm$ 0.09 \\
    \hline
    \end{tabular}
    \begin{tablenotes}
      \item NOTE--\\
      \item $^{a}$ Number of epochs detected in the speckle holography images.
      \item $^{b}$ Additional uncertainty from variability (\citealt{Gautam et al. 2018}), which for each star is less than the uncertainty of the original Blum's measurements. 
      Reported in \citet{Gautam et al. 2018}.
      \item $^{c}$ Aperture correction with a radius of $0.5''$ (see Appendix \ref{sec:calib}).
      \item $^{d}$ Predicted NIRC K mag for 4 initial calibrators obtained from \cite{Blum et al. 1996} after aperture and bandpass correction; predicted NIRC K mag for 12 verification calibrators obtained from \cite{Schodel et al. 2010} after bandpass correction. See details in Appendix \ref{sec:calib}. 
      \item $^{e}$ Measured NIRC K magnitude after initial photometric system calibration (see Appendix \ref{sec:calib}). 
      \item $^{f}$ Measured NIRC K relative magnitude after relative photometric calibration (see Appendix \ref{sec:calibrela}). 
      \item $^{g}$ Photometric uncertainty with additional variability uncertainty. 
      \item $^{h}$ Average photometric systematic zero-point uncertainty. 
      \item $^{i}$ Relative photometric uncertainty (see Appendix \ref{sec:calibrela}). The average zero-point uncertainty for relative photometry is 0.04 mag. 
    \end{tablenotes}
    }
\end{table*}

\begin{table*}
    \centering
    \tablenum{4}
    {
    \caption{Filter transformation for Sgr A*}\label{tab:filtersgra}
    \begin{tabular}{ccc}
    \hline 
    \hline 
    Name & K$^\prime_{NIRC2}$-K$_{NIRC}$ &K$_{s,NACO}$-K$_{NIRC}$ \\[2ex]
    \hline
    Sgr A* & 0.367 $^{+0.01}_{-0.02}$ & 0.275 $^{+0.01}_{-0.02}$\\[2ex]
    \hline
    \end{tabular}
    }
\end{table*}

\subsection{Relative Photometric Calibration}\label{sec:calibrela}

We perform relative photometric calibration using the secondary calibrator stars identified by \citet{Gautam et al. 2018} that are detected in the field of view of our observations. These calibrator stars, IRS 16NW, S1-17,S1-34, S1-1, S0-14 and S1-21 (see Figure \ref{fig:calibrators} Left panel), are selected to be non-variable and well-distributed in the field of view. The reference fluxes of these calibrators were obtained from the photometric system calibration described in section \ref{sec:calib} (See Table \ref{tab:bandpass}). See Figure \ref{fig:lc_sec} for the light curve with measured relative K magnitude for each of the calibrator after relative photometric calibration. The derivation of the zeropoint and uncertainties in the relative calibration is the same as for the photometric system calibration procedure (see section \ref{sec:calib}). We achieved an average uncertainty $\overline{\sigma}_{zp}$ for the relative photometric calibration of 0.04 mag in NIRC K bandpass. See Table \ref{tab:speckle1} for details of single epoch.

\section{Source analyses for Speckle Holography}\label{sec:sourceanalyses}
\subsection{Bootstrap fraction threshold and detection limit} \label{sec:lowerthreshold}

\begin{figure}
    \centering
    \includegraphics[width=4 in]{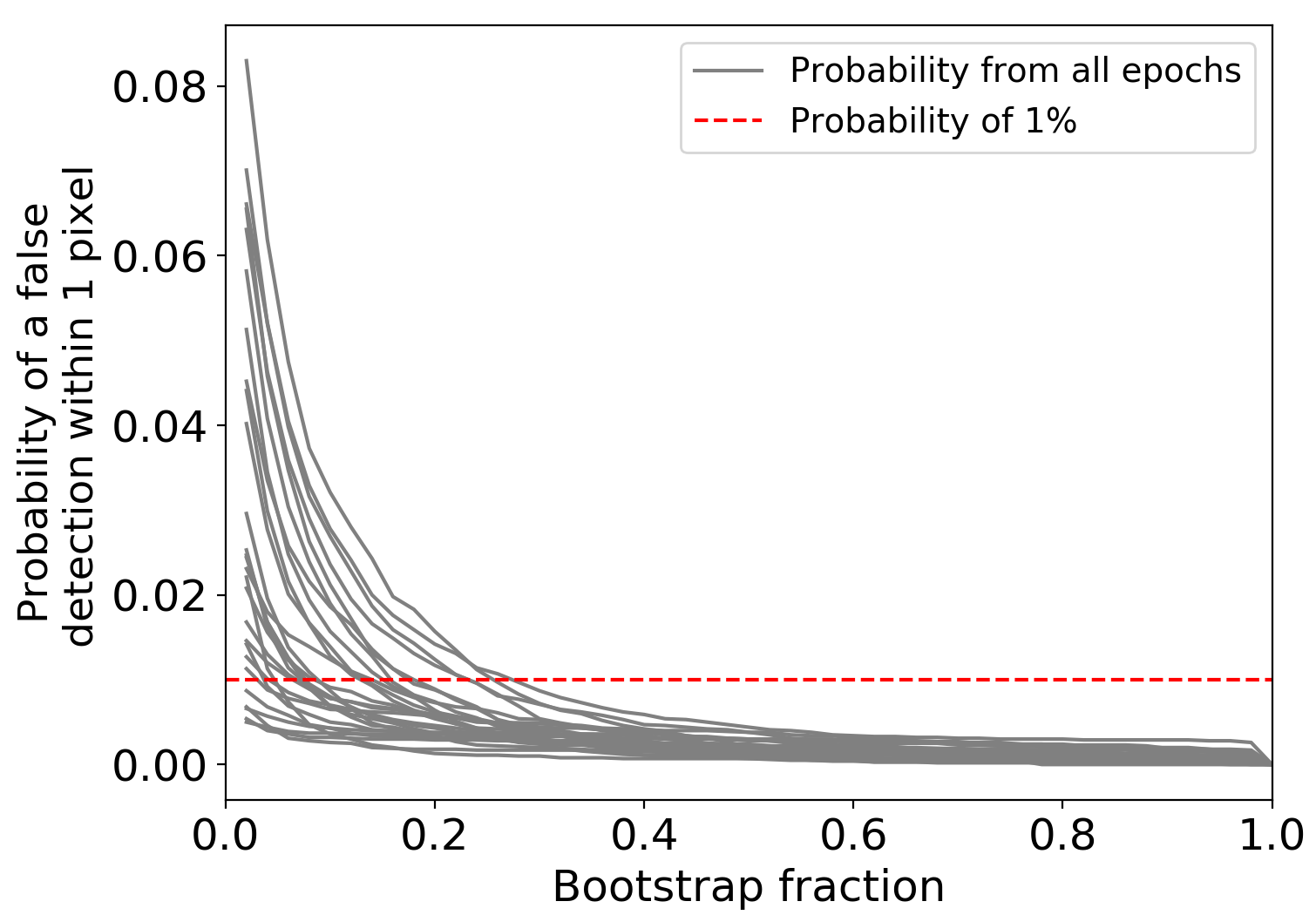}
    \caption{The probability of a false detection within 1 pixel (10 mas square) around its nominal position as a function of bootstrap fraction cut below which the detection is treated as likely spurious. Each grey line presents one epoch's probability function at all possible bootstrap fractions. The bootstrap fraction cut of a real detection in each epoch is set to be the value at which the probability of a false detection within 1 pixel is $1\%$ (red line).This ensures that all detections are real. }
    \label{fig:prob_fake}
\end{figure}

In order to define criteria for real detections, we can use the stellar photometric and astrometric information from the analysis of our extensive adaptive optics (AO) imaging data (eg., \citealt{Jia et al. 2019}), which are on the order of three magnitudes deeper than the speckle images. We define a reference set of 88 sources (real stars) in the central $2'' \times 2''$ region with $K < 17.6$ mag (deepest speckle datasets limit) and detected as the same source in at least $1/3$ of 39 AO epochs. The real sources typically have high bootstrap fractions, while the remaining detections have quite low bootstrap fractions. The bootstrap fraction of any given object is defined as the portion of bootstrap images (among overall 100 bootstraps) in which the object can be detected (see step 4 in section \ref{sec:psffitting}). We use the remaining sources to estimate the surface density of likely spurious detections in the central $2'' \times 2''$ region at each bootstrap fraction below which the detection is treated as likely spurious. Assuming a random process we can thus compute the probability of obtaining a false detection within 10 mas search radius around its nominal position, which is a function of the bootstrap fraction threshold that is applies to each epoch (see Figure \ref{fig:prob_fake}). To avoid false detections, we require that the bootstrap fraction cut of a real source detection in each epoch is obtained at which the probability of a false detection within 10 mas radius is always $< 1\%$. See Table \ref{table:chartable} for the summary of the bootstrap fraction cut.

Then the source detection limit for each epoch is determined to be the $95^{th}$ percentile of all K magnitudes in the sample which includes all sources with a bootstrap fraction higher than the threshold in the central $2'' \times 2''$ field of view. The median of the detection limit for all epochs is 16.9 mag at K, which corresponds to observed flux of 0.15 mJy and dereddened flux of 1.4 mJy respectively. See Table \ref{table:chartable} for details. 

\subsection{Comparison between Speckle Holography version 2\_1 and version 2\_2}\label{sec:v1v2com}

\begin{figure}
    \centering
    \includegraphics[width=4.5 in]{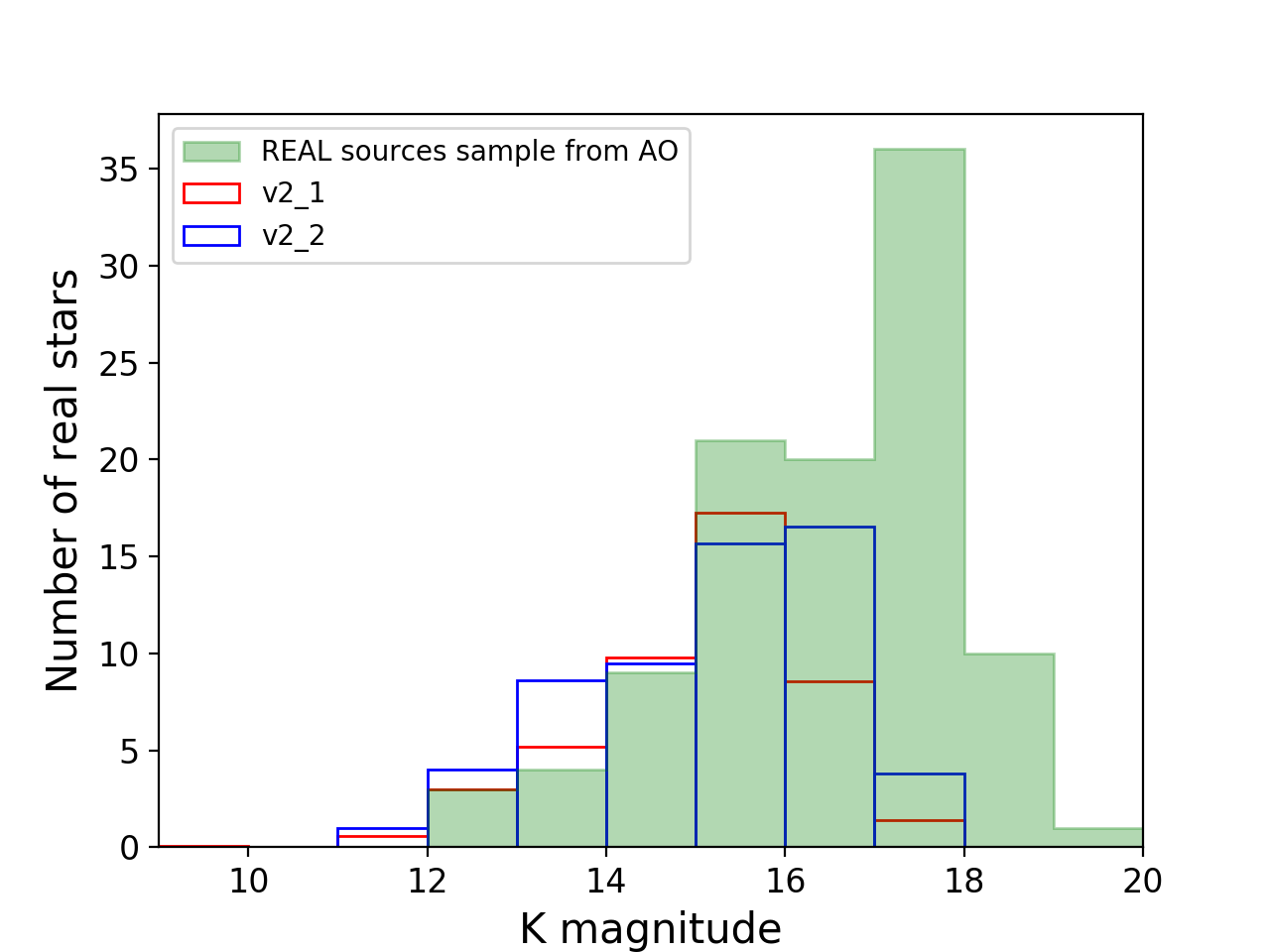}
    \caption{Comparison of the distribution of real detections based on their K-band magnitudes with the original speckle holography (version 2\_1, red) and the new implementation presented here (version 2\_2, blue). The green histogram shows the distribution of real sources sample defined in Appendix \ref{sec:lowerthreshold} with their average K$^\prime$ magnitudes from AO observations. The number of real detections for each speckle holography version plotted here is the average from 27 speckle epochs. The new analysis results in deeper detection ($\sim$ 0.4 mag deeper), and increases significantly the completeness of detection between $K = 15$ $\sim$ $17$ mag, where Sgr A* lies.}
    \label{fig:v1v2}
\end{figure}

We performed the real source analysis (see section \ref{sec:lowerthreshold}) for both speckle holography version 2\_1 and version 2\_2, and then obtained the real source lists for two datasets respectively. Here we compare speckle holography analysis version 2\_1 and version 2\_2 based on the real detection list in the central $2'' \times 2''$ region:
\begin{enumerate}[noitemsep,nolistsep]
  \item The speckle holography technique version 2\_2 results in deeper detections. The average magnitude limit for speckle data has been increased from $K = 16.5$ (version 2\_1) to $K = 16.9$ (version 2\_2). See Table \ref{tab:speckle12} for details of each epoch. See Figure \ref{fig:v1v2}. 
  \item In $>80\%$ epochs, the speckle holography technique version 2\_2 results in more real detections. The average number of real detected stars in the central $2'' \times 2''$ region has been increased from $N = 46$ (version 2\_1) to $N = 59$ (version 2\_2). See Table \ref{tab:speckle12} for details of each epoch.
  \item The version 2\_2 increases the completeness of detection between $K = 15$ $\sim$ $17$ mag, where Sgr A* lies (see Figure \ref{fig:v1v2}). 
  \item The halos around sources, especially bright ones, are reduced in the speckle holography version 2\_2 datasets. (See section \ref{sec:speckleholography} and Figure \ref{fig:data})
  \item For the speckle holography version 2\_1 datasets, we obtain the uncertainties from running StarFinder on 3 submaps (standard deviation divided by square root 3). For the version 2\_2 datasets, we calculate the uncertainties from running StarFinder on up to 100 bootstraps. This new bootstrapping technique gives mathematically more accurate uncertainties. 
\end{enumerate}
\vspace{10mm}

\begin{deluxetable*}{lcccrrrcrcrc}[t]
\tabletypesize{\footnotesize}
\tablenum{5}
\tablecolumns{6}
\tablewidth{0pt}
\tablecaption{Comparison between Speckle Holography version 2\_1 and version 2\_2
\label{tab:speckle12}}

\tablehead{
\multicolumn{2}{c}{Date} & \multicolumn{2}{c}{$K_{lim}$ (mag)} & \multicolumn{2}{c}{$N_{Real Stars}$} & \multicolumn{2}{c}{$N_{pix}$ $^{b}$} & \multicolumn{2}{c}{Max Frames} & \multicolumn{2}{c}{$N_{ref}$ $^{c}$} \\ \cline{1-12} \colhead{(U.T.)} & \colhead{(Decimal)} & \colhead{Value$^{d}$} & \colhead{$\Delta m^{e}$} & \colhead{Value} & \colhead{Ratio} & \colhead{Value} & \colhead{Ratio} & \colhead{Value} & \colhead{Ratio} & \colhead{Value} & \colhead{Ratio}
}

\startdata
1995 Jun 9-12 & 1995.439 & 17.0& 1.15 & 41 &1.21& 108042 & 0.90 & 5286 & 1.24 & 19 & 0.95 \\
1996 Jun 26-27 & 1996.485 & 15.8 & 0.36 & 49 & 1.40&  82505 & 0.79 & 2336 & 0.52 & 22 & 1.10 \\
1997 May 14 & 1997.367 & 16.8 & 0.48 & 51 & 1.21&  92467 & 0.74 & 3486 & 2.99 & 25 & 0.83  \\
1998 Apr 2-3 & 1998.251 & 15.8 & 0.25 & 39 & 1.00 & 95816 & 0.78 & 1730 & 0.83 & 24 & 0.81  \\
1998 May 14-15 & 1998.366 & 16.8 & -0.02 & 45 &0.92 &  102328 & 0.82 & 7685 & 0.77 & 24 & 0.89  \\
1998 Jul 3-5 & 1998.505 & 16.4 & 0.57 & 43 & 1.08 & 116557 & 0.83 & 2053 & 0.81 & 24 & 0.83  \\
1998 Aug 4-6 & 1998.590 & 17.1 & 0.16 & 47 & 0.94 & 109269 & N/A$^{f}$ & 11047 & 0.46 & 23 & 0.77  \\
1998 Oct 9,11 & 1998.771 & 16.6 & 0.41 & 45 & 1.22 & 97215 & 0.81 & 2015 & 0.87 & 24 & 0.80  \\
1999 May 2-4 & 1999.333 & 17.2 & 0.10 & 52 & 0.96 & 107882 & 0.77 & 9427 & 0.96 & 22 & 0.81  \\
1999 Jul 24-25 & 1999.559 & 17.4 & 0.73 & 54 & 1.02 & 100567 & 0.76 & 5776 & 0.99 & 23 & 0.79  \\
2000 Apr 21 & 2000.305 & 15.7 & 0.13 & 56 &1.81 &  96248 & 0.78 & 662 & 0.21 & 21 & 0.84  \\
2000 May 19-20 & 2000.381 & 17.5 & 0.39 & 55 &0.89 &  96853 & 0.80 & 15591 & 0.98 & 23 & 0.79  \\
2000 Jul 19-20 & 2000.584 & 17.0 & 0.32 & 63 &1.29 &  86452 & 0.78 & 10678 & 0.98 & 23 & 0.82  \\
2000 Oct 18 & 2000.797 & 16.2 & 0.51 & 52 & 1.73 & 82315 & 0.84 & 2247 & 0.88 & 17 & 0.74  \\
2001 May 7-9 & 2001.351 & 17.2 & 0.53 & 64 & 1.28 & 85028 & 0.91 & 6678 & 0.85 & 21 & 0.84  \\
2001 Jul 28-29 & 2001.572 & 17.4 & 0.22 & 74 &1.21 &  96872 & 0.78 & 6654 & 0.99 & 23 & 0.85 \\
2002 Apr 23-24 & 2002.309 & 17.5 & 0.65 & 74 & 1.30 & 96953 & 0.79 & 13469 & 0.98 & 23 & 0.82 \\
2002 May 23-24 & 2002.391 & 17.6 & 0.51 & 72 & 1.22 & 98552 & 0.83 & 11860 & 0.99 & 21 & 0.78  \\
2002 Jul 19-20 & 2002.547 & 16.8 & 0.59 & 69 & 1.73 & 99994 & 0.79 & 4192 & 0.72 & 22 & 0.81 \\
2003 Apr 21-22 & 2003.303 & 16.4 & 0.32 & 58 & 1.49 & 90963 & 0.93 & 3715 & 0.89 & 23 & 0.96  \\
2003 Jul 22-23 & 2003.554 & 16.8 & 0.25 & 65 & 1.41& 87265 & 0.79 & 2914 & 0.96 & 24 & 0.86 \\
2003 Sep 7-8 & 2003.682& 17.1 & 0.60 & 74 & 1.57 & 95367 & 0.79 & 6324 & 1.00 & 20 & 0.77 \\
2004 Apr 29-30 & 2004.327 & 16.8 & 0.15 & 58 & 1.07 & 125423 & 0.99 & 6212 & 0.51 & 26 & 1.00 \\
2004 Jul 25-26 & 2004.564 & 17.4 & 0.48 & 80 & 1.45 & 99819 & 0.78 & 13085 & 0.99 & 22 & 0.85  \\
2004 Aug 29 & 2004.660 & 16.8 & 0.60 & 63 & 1.54 & 96172 & 0.96 & 2299 & 0.75 & 25 & 0.93 \\
2005 Apr 24-25 & 2005.312 & 17.1 & 0.24 & 70 & 1.46& 105715 & 0.81 & 9644 & 0.88 & 24 & 0.89 \\
2005 Jul 26-27 & 2005.566 & 16.8 & 0.81 & 84 &2.33 & 108360 & 0.79 & 5642 & 0.96 & 23 & 0.92 \\
\enddata
\tablecomments{\\
$^{a}$ $K_{lim}$ is the magnitude that corresponds to the 95$^{th}$ percentile of all $K$ magnitudes in the sample of real Stars in the central $2'' \times 2''$ region (see section \ref{sec:lowerthreshold}). \\
$^{b}$ $N_{pix}$ refers to the number of pixels in a given image that meet a $.8$ of maximum frames used criteria. \\
$^{c}$ $N_{ref}$ refers to the number of reference stars used to align the epoch of data.  \\
$^{d}$ All values given in the table are the results of version 2\_2.\\
$^{e}$ $\Delta m = K_{lim, 2\_2} - K_{lim, 2\_1}$, is the detection limit difference between the version 2\_2 and the version 2\_1. The average magnitude limit has been increased from $K = 16.5$ (version 2\_1) to $K = 16.5$ (version 2\_2).\\
$^{f}$ The number of pixels in Aug 1998 is removed due to an artifact in the old holography image.
}
\end{deluxetable*}

\end{document}